\documentclass[pra,showpacs,aps,amssymb,12pt]{revtex4}
\usepackage{graphicx}
\usepackage{amsmath}
\usepackage{feynmf}

\newcommand{\newc}{\newcommand}
\newc{\beq}{\begin{equation}}
\newc{\enq}{\end{equation}}
\newc{\bea}{\begin{eqnarray}}
\newc{\ena}{\end{eqnarray}}
\newc{\D}{\displaystyle}
\newc{\noi}{\noindent}
\newc{\cA}{{\cal A}}
\newc{\cB}{{\cal B}}
\newc{\cC}{{\cal C}}
\newc{\cD}{{\cal D}}
\newc{\cE}{{\cal E}}
\newc{\cN}{{\cal N}}
\newc{\cF}{{\cal F}}
\newc{\cG}{{\cal G}}
\newc{\cH}{{\cal H}}
\newc{\cI}{{\cal I}}
\newc{\cJ}{{\cal J}}
\newc{\cK}{{\cal K}}
\newc{\cL}{{\cal L}}
\newc{\cM}{{\mathbf {M}}}
\newc{\cO}{{\cal O}}
\newc{\cP}{{\cal P}}
\newc{\cPr}{{\cal P}}
\newc{\cQ}{{\cal Q}}
\newc{\cR}{{\cal R}}
\newc{\cS}{{\cal S}}
\newc{\cT}{{\cal T}}
\newc{\mG}{{\mathbf G}}
\newc{\Gg}{{\{\mG\cG\}}}
\newc{\GG}{{\{\mG\mG\}}}
\newc{\Gbarra}{{\{\mG\!\!\!\!\slash\,\}}}
\newc{\gbarra}{{\{\cG\!\!\!\!\slash\,\}}}
\newc{\fbarra}{{f_k}}
\newc{\mF}{{\mathbf F}}
\newc{\mi}{i}
\newc{\sen}{{\rm sen\;}}
\newc{\ra}{\rightarrow}
\newc{\rt}{\right}
\newc{\lt}{\left}
\newc{\sqp}{\phi}
\newc{\cp}{\varphi}
\newc{\csp}{\phi_{cl}}
\newc{\ccp}{\varphi_{cl}}
\newc{\osp}{\overline {\phi}}
\newc{\ocp}{\overline {\varphi}}
\newc{\cj}{{\it j}}
\newc{\sj}{{\mathbf J}}
\newc{\cpmm}{{\varphi_>}}
\newc{\cpm}{{\varphi_<}}
\newc{\spmm}{{\phi_>}}
\newc{\spm}{{\phi_<}}
\newc{\cjmm}{{\cj_>}}
\newc{\cjm}{{\cj_<}}
\newc{\sjmm}{{\sj_>}}
\newc{\sjm}{{\sj_<}}
\newc{\etamm}{{\eta_>}}
\newc{\etam}{{\eta_<}}
\newc{\tcp}{{\tilde {\cp}}}
\newc{\tsp}{{\tilde {\sqp}}}
\newc{\nn}{\nonumber}
\newc{\ob}{\overbrace}
\newc{\ub}{\underbrace}
\newc{\veco}{\Lambda \hat \Omega}
\newc{\bfK}{{\mathbf K}}
\newc{\bfk}{{\mathbf k}}
\newc{\bfq}{{\mathbf q}}
\newc{\bfQ}{{\mathbf Q}}
\newc{\bfP}{{\mathbf P}}
\newc{\bfD}{{\mathbf D}}
\newc{\bfp}{{\mathbf p}}
\newc{\la}{\Lambda}
\newc{\bla}{\overline\Lambda}
\newc{\Ci}{{\rm Ci}}
\newc{\Si}{{\rm Si}}
\newc{\Ei}{{\rm Ei}}
\newc{\poi}{\phantom . \noi}

\begin{document}

\title{Renormalization group study of damping in nonequilibrium field theory}
\author{Juan Zanella\footnote{Email address zanellaj@df.uba.ar}
and Esteban Calzetta\footnote{Email address calzetta@df.uba.ar}}

\affiliation{CONICET and Departamento de F\'{\i}sica, Universidad
de Buenos Aires\\ Ciudad Universitaria, 1428 Buenos Aires,
Argentina}

\setlength{\baselineskip}{16pt}

\begin{abstract}
In this paper we shall study whether dissipation in a
$\lambda\varphi^{4}$ may be described, in the long wavelength, low
frequency limit, with a simple Ohmic term $\kappa\dot{\varphi}$,
as it is usually done, for example, in studies of defect formation
in nonequilibrium phase transitions. We shall obtain an effective
theory for the long wavelength modes through the coarse graining
of shorter wavelengths. We shall implement this coarse graining by
iterating a Wilsonian renormalization group transformation, where
infinitesimal momentum shells are coarse-grained one at a time, on
the influence action describing the dissipative dynamics of the
long wavelength modes. To the best of our knowledge, this is the
first application of the nonequilibrium renormalization group to
the calculation of a damping coefficient in quantum field theory.
\end{abstract}

\pacs{11.10.-z, 11.10.Hi, 05.40.Ca}

\maketitle

\section{Introduction}

\label{sec:Introduction} In this paper we shall study whether
dissipation in a $\lambda\varphi^{4}$ theory may be described, in
the long wavelength, small frequency limit, with the addition of a
simple Ohmic term $\kappa\dot{\varphi}$, as it is usually done,
for example, in studies of defect formation in nonequilibrium
phase transitions. We shall obtain an effective theory for the
long wavelength modes through the coarse graining of shorter
wavelengths. We shall implement this coarse graining by iterating
a Wilsonian renormalization group (RG) transformation
\cite{WiKo74,WH,Polchinski}, where infinitesimal momentum shells
are coarse-grained one at a time, of the influence action
describing the dissipative dynamics of the long wavelength modes.
To the best of our knowledge, this is the first application of the
nonequilibrium RG to the calculation of a transport coefficient in
quantum field theory.

Understanding damping is one of the main goals of nonequilibrium
quantum field theory, particularly with respect to (probably) its
main applications, namely the generation of primordial
fluctuations during Inflation \cite{LoNa05}, the modelling of
reheating after Inflation \cite{BTW06}, and the formation of
defects in  nonequilibrium phase transitions \cite{kzm}. There is
a tension between the needs of model builders, who,
understandably, seek simple and model-robust solutions, and first
principles calculations, whose generic result is that damping in
field theory is a complicated, non Markovian phenomenon
\cite{CH89,Law05,BERA01,BGR98}. A clear depiction of this tension
can be seen in the contrast between the early models of reheating,
based on a linear damping of the inflaton field, and the now
accepted picture of preheating, where the main decay mechanism for
the inflaton is the parametric amplification of matter fields.
This phenomenon is exponentially suppressed in perturbative
calculations \cite{CH97}.

The physical mechanism for damping in the long wavelength sector
is the interaction with shorter wavelength modes. Damping is a
feature of the effective theory where the shorter modes have been
coarse-grained away \cite{CHM}. Since this operation will leave
the long wavelength modes in a mixed state, the natural
description of the relevant sector is in terms of a density matrix
\cite{Pol06}, and the natural action functional encoding the
effective dynamics is the Feynman-Vernon Influence action
\cite{if}. Now suppose we are given the influence action when all
modes $k>\Lambda$ have been coarse-grained away, and we wish to
further coarse grain the modes in the range $\Lambda\ge k >
k_{0}$. A straightforward perturbative functional integration (for
example, to two-loops accuracy) is likely to miss the most
important effects. However, if we split the desired range into
shells of infinitesimal thickness $\delta s$, then each shell can
be integrated out exactly within one-loop accuracy, because each
loop integration brings a factor of $\delta s$. Adding a change of
units after each integration, we transform the shell
coarse-graining into a RG flow in the space of influence actions.
Because we shall not assume equilibrium conditions, this may be
called the nonequilibrium renormalization group. Our goal is to
study the flow of the damping parameter $\kappa$ in the infrared
limit.

The nonequilibrium renormalization group should not be confused
with the so-called dynamical RG , which is concerned with the long
term behavior of solutions to evolution equations
\cite{DyRG1,DyRG2,DyRG3,DyRG4,Boyanovsky:Vega:Simionato}. The
nonequilibrium RG, on the other hand, is closely related to the
Martin, Siggia and Rose formalism \cite{Martin:Siggia:Rose}. The
equivalence between these two latter approaches has been discussed
elsewhere \cite{us,Cooper:Rose}.

It is also important to stress the difference between our approach
and the RG applied to dynamic phenomena, as in
\cite{Ma,alternative:rg:dynamics1,alternative:rg:dynamics2,
alternative:rg:dynamics3,alternative:rg:dynamics4,alternative:rg:dynamics5}.
In these applications, a dissipative dynamics is assumed from
scratch, as well as a phenomenological noise term to induce the
fluctuations to be coarse grained away. In our application,
instead, we shall assume that the initial point of the RG
trajectory corresponds to a nondissipative theory. The source of
noise will be just the quantum fluctuations in the short
wavelength modes, as well as any statistical noise coming from the
initial density matrix (see below). The issue is whether the
renormalization group flow alone brings in dissipation and noise.

It is important to stress two basic differences between the
nonequilibrium and equilibrium renormalization groups
\cite{Attanasio1,Attanasio2}. The influence action may be regarded
as an action for a theory defined on a ``closed time path'' (CTP)
composed of two branches \cite{ctp}. The first branch goes from
the initial time $t=0$ to a later time $t=T$ when the relevant
observations will be performed; that is why we need the density
matrix at $T$. The second branch returns from $T$ to $0$. Thus
each physical degree of freedom on the first branch acquires a
twin on the second branch -we say the number of degrees of freedom
is doubled. The influence action is not just a combination of the
usual actions for each branch, but also admits direct couplings
across the branches. The damping constant $\kappa$ is associated
to one of these ``mixed'' terms. Therefore, the structure of the
influence action (from now on, CTP action, to emphasize this
feature) is much more complex than the usual Euclidean or
``IN-OUT'' action.

The second fundamental difference is the presence of the parameter
$T$ itself. In nonequilibrium evolution, it is important to
specify the time scale over which we shall observe the system
\cite{Pomeau:Arnold}. The CTP action contains this physical time
scale $T$. From the point of view of the RG, this adds one more
dimensional parameter to the theory, much as an external field in
the Ising model. Physically, because time integrations are
restricted to the interval $\left[0,T\right]$, energy conservation
does not hold at each vertex. This is of paramount importance
regarding damping.

The Wilsonian RG for equilibrium quantum fields in the imaginary
time formalism was studied in Ref. \cite{Liao:Polonyi:Xu}, and in
the real time formalism in Ref. \cite{Attanasio1}. The RG for the
CTP effective action (obtained by taking the limit
$T\rightarrow\infty$) was studied by Dalvit and Mazzitelli
\cite{Dalvit:Mazzitelli}; see also \cite{CHM} and \cite{Pol06}.
Unlike those works, we focus on the dissipation and noise features
of the effective dynamics, rather than in the running of the
effective potential. For an application of similar ideas in a
different field see \cite{Gezzi}.

We shall now describe the basic assumptions underlying our work.

In formulating a nonequilibrium RG, we must deal with the fact
that the CTP action may have an arbitrary functional dependence on
the fields and be nonlocal both in time and space. In principle,
one can define an exact RG transformation
\cite{WH,Wetterich,Dalvit:Mazzitelli}, where all three functional
dependencies are left open. However, the resulting formalism is
too complex to be of practical use
\cite{Dalvit:Mazzitelli,Morris}. At the same time, one must beware
of restricting the form of the action to the point of leaving out
an important process. An example is, if we were to assume that the
CTP action is the difference of an action functional for each
branch of the CTP, we shall miss the damping and noise terms. This
is allowed, of course, if one interest is, for example, the
running of the effective potential, but it would be disastrous to
our present concern.

In our case, we will assume that the initial condition for the RG
trajectory corresponds to a massless $\lambda\varphi^{4}$ theory,
with no noise or dissipation. This introduces a small parameter
$\lambda$ in the model, and we will consider only the class of
action functionals which may be generated from this initial
condition to order $\lambda^{2}$. This means we shall include
quadratic, quartic and six-field interactions. We shall allow the
interaction terms to be nonlocal in time and in space, to the
extend demanded by closure of the RG transformation. There are
further constraints coming from the CTP boundary conditions; these
will be described below.

We shall work in three spatial dimensions. The structure of the
nonequilibrium RG becomes much richer below this critical
dimension, since there several fixed points (over and above the
nontrivial fixed point of the equilibrium RG). These will be
described in a separate publication. However, for the applications
listed at the beginning, three spatial dimensions is the most
relevant choice. We shall assume that the initial density matrix
decomposes into an independent matrix for each mode. The formalism
may be developed with great generality, but to obtain definite
results we will focus in the case were the initial density matrix
corresponds to a $free$ field at finite temperature. This is a
nonequilibrium initial condition for the $interacting$ field
\cite{Baacke:Boyanovsky:Vega}.

We are mostly interested in the case of small $T$ (that is,
$\Lambda T\approx 1$), since for large $T$ we expect the
nonequilibrium RG to converge to the usual one (we shall return to
this point below). This means that in computing Feynman graphs, it
is allowed to use propagators pertaining to the initial density
matrix and couplings, as any drift term will bring additional
powers of $\lambda$. To extend the nonequilibrium RG to a larger
$T$ range (for example, to study thermalization within this
approach) a fully self-consistent approach is necessary
\cite{self:consistent:approach} in the spirit of the so-called
``environmentally friendly'' RG \cite{SO}.

The paper is organized as follows. In the next two sections we
provide an introduction to the CTP formalism, thus fixing our
notation, and introduce the parametrization of the CTP action and
of the initial density matrix. As we have remarked, the form of
the CTP action is chosen to enforce closure of the RG
transformation, within the constraints imposed by CTP boundary
conditions. In Sec. \ref{section the RG transformation} we define
the RG transformation. This is composed of two stages, first the
elimination of one shell in momentum space, thus lowering the
cutoff to $(1-\delta s)\Lambda$, and then the rescaling of
momenta, times and fields to restore the value of the cutoff and
the coefficients of the kinetic terms in the CTP action. We will
show that, in spite of rescaling, the parameter $T$ may be kept as
a RG invariant. We then translate the change in the CTP action
into a dynamic equation for the RG flow in parameter space. In
Sec. \ref{section the RG equations to order lambda2} we write down
these RG equations to order $\lambda^{2}$ and discuss their
structure. In Sec. \ref{section kappa and nu} we discuss the flow
of the damping parameter $\kappa$ and of the noise kernel $\nu$ in
the regime outlined above.

We conclude the paper with some brief final remarks. We have
concentrated most technical details into appendixes.

\section{Open systems, CTP and RG}

The core of the RG transformation is the elimination of selected
degrees of freedom. When we are only concerned with  the lower
wave number sector of a field, we can carry out explicitly the
integration over the higher wave number modes in the density
matrix. As a result, these modes are eliminated from the
description. This partial integration will return a description
for the lower wave number modes only. The influence of the higher
modes is incorporated in the parameters which define the effective
action for the surviving modes. The procedure may be seen as a
straightforward application of the Feynman-Vernon influence
functional techniques for open systems. In the problem at hand,
the long modes are regarded as the system and the short ones as
the environment.

We first review briefly the ideas behind the Feynman-Vernon
formalism. We will deal with a system composed of two parts, which
we call the relevant system, or simply the system, and the
environment.

\subsection{Open systems and Feynman-Vernon formalism
\label{section open systems and Feynman-Vernon formalism}}

Let $\cp$ be the field variables of the whole system, and $S$ its
action. Then, its density matrix $\rho$ admits the following
representation \bea \rho[\cp^+, \cp^-, T] &=&
\lt<\cp^+\Big|U(0, T)\,\rho \,U(T, 0)\Big|\cp^-\rt> \nn \\ \nn \\
\label{density matrix 1} &=&
{\mathrel{\mathop{\lower3.0ex\hbox{\kern-.190em $_{\cp^+(T) =
\cp^+, \;  \cp^-(T) = \cp^-}$}}}} \!\! \!\! \!\! \!\! \!\! \!\!
\!\! \!\! \!\! \!\! \!\! \!\!\! \!\! \!\! \!\! \!\! \!\! \!\!\
\!\! \!\! \!\! \!\! \!\! \D \int \cD \cp^+ \cD \cp^- \; \exp i
\lt(S[\cp^+]-S[\cp^-]\rt) \, \rho[\cp^+(0), \cp^-(0), 0] .\ena
This expression says how to obtain the matrix elements of the
density matrix operator in the Schr{\"o}dinger representation at
time $T$. The generating functional for the expectation values is
obtained by adding current terms, and by closing the path of
integration to take the trace \bea \label{Z whole system} Z[J^+,
J^-] ={\mathrel{\mathop{\lower3.0ex\hbox{\kern-.190em $_{\cp^+(T)
= \cp^-(T)}$}}}} \!\! \!\! \!\! \!\! \!\! \!\! \!\! \!\!\ \!\!
\!\! \!\! \!\! \!\!\!\! \!\! \D \int \cD \cp^+ \cD \cp^- \; \exp i
\lt\{S[\cp^+]-S[\cp^-] + \D\int (J^+ \cp^+ + J^- \cp^-)\rt\} \,
\rho[\cp^+(0), \cp^-(0), 0]\,. \ena Derivatives of $Z$ with
respect to the currents give expectation values of products of the
fields (with a time ordering that depends on which derivatives
have been taken). Equation (\ref{Z whole system}) can be thought
as an integral over single histories defined on a closed time
path. This path has a first branch from $0$ to $T$, where the
history takes the values $\cp^+(t)$, and a second branch from $T$
to $0$, where the history takes the values $\cp^-(t)$. The CTP
condition \bea \cp^+(T) = \cp^-(T) \label{CTP condition} \ena
implies that each history is a continuous function of the time
along the path. The object \bea S_{\rm CTP}[\cp^+, \cp^-] =
S[\cp^+] - S[\cp^-] \label{def CTP action} \ena is called the CTP
action. In general it can be written \bea \label{density matrix 2}
\rho[\cp^+, \cp^-, T] &=&
{\mathrel{\mathop{\lower3.0ex\hbox{\kern-.190em $_{\cp^+(T) =
\cp^+, \;  \cp^-(T) = \cp^-}$}}}} \!\! \!\! \!\! \!\! \!\! \!\!
\!\! \!\! \!\! \!\! \!\! \!\!\! \!\! \!\! \!\! \!\! \!\! \!\!\
\!\! \!\! \!\! \!\! \!\! \D \int \cD \cp^+ \cD \cp^- \; \exp\lt(i
S_{\rm CTP}[\cp^+, \cp^-]\rt) \, \rho[\cp^+(0), \cp^-(0), 0] .\ena

Now, we make the division between relevant system and environment.
Suppose that the whole system consists on two kind of variables
$\cp_<$ and $\cp_>$, being $\cp_<$ the  system variables, and
$\cp_>$ the environment variables. Depending on the context, this
could be a division between system and bath fields in the
thermodynamical sense, or between different particle fields, or,
as will be in our case, two different sectors of the same field:
the lower and the higher wave number sectors.

The whole system is described using the generating functional
associated with the full density matrix, which depends on both
system and environment variables. However, if we only wish to
compute expectation values for the system observables, we can use
the generating functional associated with the reduced density
matrix  that is obtained after taking the trace with respect to
the environment variables. To do this, suppose that the CTP action
of the system + environment is $S_{\rm CTP}[\cp^+, \cp^-]$, then
write it as a functional of the two pairs of fields, $(\cp^+_<,
\cp^-_<)$ and $(\cp^+_>,\cp^-_>)$, for the system and the
environment, respectively,  \bea \label{division} S_{\rm
CTP}[\cp^+, \cp^-] = S_{\rm CTP}[\cp^+_<, \cp^-_<]+ \Delta
S[\cp^+_<, \cp^-_<, \cp^+_>, \cp^-_>]. \ena For future
convenience, the same functional $S_{\rm CTP}$  appears in both
members; this is just the definition of $\Delta S$. The functional
$\Delta S$ contains the information about the environment itself
and about the interaction with the system. Let us further assume
that at $t= 0$ system and environment are uncorrelated, \bea
\rho(0) = \rho_s(0) \otimes \rho_e(0), \label{rho uncorrelated}
\ena where $\rho_s(0)$ ($\rho_e(0)$) refers to the initial density
matrix of the proper system (environment). After, replacing
(\ref{division}) in the Eq. (\ref{density matrix 2}) and
integrating out the environment variables $\cp_>^\pm$,  we get an
evolution law for the so called reduced density matrix \bea
\rho_r[\cp_<^+, \cp_<^-, T] =
{\mathrel{\mathop{\lower3.0ex\hbox{\kern-.190em $_{\cp_<^+(T) =
\cp_<^+, \;  \cp_<^-(T) = \cp_<^-}$}}}} \!\! \!\! \!\! \!\! \!\!
\!\! \!\! \!\! \!\! \!\! \!\! \!\!\! \!\! \!\! \!\! \!\! \!\!
\!\!\ \!\! \!\! \!\! \!\! \!\! \D \int \cD \cp_<^+ \cD \cp_<^- \;
\exp i \lt(S_{\rm CTP}[\cp_<^+, \cp_<^-] + S_{IF}[\cp_<^+,
\cp_<^-]\rt)\; \rho_s[\cp_<^+(0), \cp_<^-(0), 0], \ena where the
influence action $S_{IF}$ is given by \bea \label{SIF} e^{i
S_{IF}[\cp_<^+, \cp_<^-]} =
{\mathrel{\mathop{\lower3.0ex\hbox{\kern-.190em $_{\cp_>^+(T) =
\cp_>^-(T)}$}}}} \!\! \!\! \!\! \!\! \!\! \!\! \!\! \!\!\ \!\!
\!\! \!\! \!\! \!\!\!\! \!\! \D \int \cD \cp_>^+ \cD \cp_>^- \;
\exp i \lt(\Delta S[\cp_<^+, \cp_<^-, \cp_>^+, \cp_>^-]\rt) \;
\rho_e[\cp_>^+(0), \cp_>^-(0), 0]. \ena All the influence of the
environment on the system is encoded into $S_{IF}$. The
elimination of the fields $\cp_>$ takes the initial CTP action,
$S_{\rm CTP}$, and returns $S_{\rm CTP<}$, the coarse-grained CTP
action for the proper system: \bea \label{CGA} S_{\rm
CTP<}[\cp^+_<, \cp^-_<] = S_{\rm CTP}[\cp_<^+,\cp_<^-] +
S_{IF}[\cp_<^+, \cp_<^-],\ena which is not necessarily the
difference of some functional evaluated in each branch, $\tilde
S[\cp_<^+] - \tilde S[\cp_<^-]$ for some $\tilde S$ as in Eq.
(\ref{def CTP action}), but usually a more complex functional. It
can entangle the two branches. CTP actions of this general type
satisfy certain properties, namely \bea S_{\rm CTP}[\cp, \cp] = 0,
\label{propiedad 1}\ena \bea S_{\rm CTP}[\cp^-, \cp^+] = -S_{\rm
CTP}[\cp^+, \cp^-]^*. \label{propiedad 2} \ena Thus, if we
introduce new variables \bea \sqp &=& \cp^+ - \cp^-, \\ \cp &=&
\cp^+ + \cp^-,\ena the expansion of $S_{\rm CTP}$ in powers of
$\sqp$, with functional coefficients $\sigma_n[\cp; t_1, \dots,
t_n]$ depending on $\cp$ and $n$ time variables \bea S_{\rm
CTP}[\sqp, \cp] = \sum_{n\ge1} \int dt_1 \dots dt_n \;
\sigma_n[\cp; t_1, \dots, t_n] \; \sqp(t_1) \dots \sqp(t_n), \ena
should start from $n = 1$, and all the odd (even) terms should be
real (imaginary). In what follows the fields $\sqp$ and $\cp$ will
be used instead of $\cp^+$ and $\cp^-$; the CTP condition
(\ref{CTP condition}) is now expressed as \bea \sqp(T) = 0
\label{CTP condition 2}. \ena

The application of this formalism to the case where the system and
the environment are two sectors of the same scalar field is
presented in Appendix \ref{section scalar field considered}.

\section{Parametrization of the CTP action and of the
initial density matrix. \label{section the action}}

{In this section we introduce the parametrization of the action
and of the initial density matrix. The parametrization of the
action will be quite general. In Sec. \ref{section the RG
transformation} we will define the RG transformation for this
general class of actions, without any further assumption regarding
the initial condition at cutoff $\Lambda$. In  Sec. \ref{section
the RG equations to order lambda2} the initial condition will be
fixed and an approximation scheme introduced. This will allow us
to reduce the parameters to be considered to a finite number of
couplings with simpler dependencies on the momenta and time
variables.

We will use natural units in which $\Lambda = 1$.} To compute the
RG transformation we will assume that the action is \bea
&\label{the action 3} S[\sqp, \cp] = S_0[\sqp, \cp] + S_{\rm
int}[\sqp, \cp],& \ena where the free action is \bea \nn &
S_0[\sqp, \cp] =  \D\int_{0}^{T} dt \, \D\int_\Lambda d^d\!k
\;\;\Bigg[ \D\frac{1}{2}  \dot\sqp(\bfk, t) \; \dot\cp(-\bfk, t) -
\D\frac{1}{2} \sqp(\bfk, t) \lt(k^2 +
m^2\rt) \cp(-\bfk, t)  & \\
\nn \\ \nn & -\kappa \;  \sqp(\bfk, t) \dot\cp(-\bfk, t) +
\D\frac{i}2 \nu \; \sqp(\bfk, t)  \sqp(-\bfk, t)\Bigg],&
\label{the action 2}  \ena and the interaction part  \bea S_{\rm
int}[\sqp, \cp] = \sum_{{\rm even} \, n \,\ge 2} S_n[\phi, \cp].
\label{expansion Sint} \ena For $n = 2$ \bea &S_2[\sqp, \cp] =
\D\int_{0}^{T} \!\! dt_1 \D\int_{0}^{T} \!\! dt_2 \D\int_\Lambda
d^d\!k \; \Big[v_{21}(k;
t_1, t_2) \; \sqp(\bfk, t_1) \, \cp(-\bfk, t_2) & \nn \\ \nn \\
\label{S2} &+ \, i \, v_{22}(k; t_1, t_2) \; \sqp(\bfk, t_1) \,
\sqp(-\bfk, t_2) \Big],& \ena and for $n > 2$ \bea &S_n[\phi, \cp]
= \Omega_d^{1-\frac{n}2} \D \lt(\prod_{j=1}^n \int_{0}^{T} dt_j
\D\int_\Lambda d^d\!k_j\rt) \, \delta^d\lt(\sum_{l = 1}^n \bfk_l\rt)  & \nn \\
\nn \\ \label{Sn} &  \times \D\sum_{m=1}^n i^{[1+(-1)^m]/2}\;\;
v_{nm}(\{\bfk\}; \{t\}) \, \; \sqp^m \; \cp^{n-m} .&\ena  {The
factor $\Omega_d^{1-n/2}$, where $\Omega_d$ is the area of the
unit sphere in $d$ dimensions, is introduced for further
convenience. The subscript $\Lambda$ in the integral symbols means
that the integrations are restricted to $|k_i|\le \Lambda = 1$.}
Moreover, we have defined \bea v_{nm}(\{\bfk\}; \{t\}) &=&
v_{nm}(\bfk_1,\dots, \bfk_m, \dots, \bfk_n; \; t_1, \dots, t_m,
\dots, t_n) \hspace{0.7cm} ({\rm with} \;\;\; 1 \le m \le n), \nn
\\  \sqp^m \;\, \cp^{n-m} &=& \sqp(\bfk_1, t_1) \dots \,
\sqp(\bfk_m, t_m) \;\, \cp(\bfk_{m+1}, t_{m+1}) \dots \,
\cp(\bfk_n, t_n). \ena {With these definitions, both the fields
and the couplings are dimensionless. Note that the CTP action
associated with the usual classical action for a massless $\lambda
\sqp^4$ is \bea &\nn S[\sqp, \cp] = \D\int_{0}^{T} dt \,
\D\int_\Lambda d^d\!k \; \D\frac{1}{2} \bigg[\dot\sqp(\bfk, t) \;
\dot\cp(-\bfk, t) - k^2 \sqp(\bfk, t)
\cp(-\bfk, t) \bigg]  & \\
\nn \\ \nn &-\D\frac\lambda{48}\D\int_{0}^{T} dt \, \D\int_\Lambda
\D\frac{d^d\!k_1 \dots d^d\!k_4}{(2\pi)^{d}}  \;
\delta^d\lt(\sum_{l=1}^4 \bfk_l\rt) \; \Big[\sqp(\bfk_1, t)
\cp(\bfk_2,t) \cp(\bfk_3,t) \cp(\bfk_4,t)  &
\\ \nn \\ \label{the action 1} &+\sqp(\bfk_1, t) \sqp(\bfk_2,t) \sqp(\bfk_3,t)
\cp(\bfk_4,t)\Big].& \ena Hence, in the notation just introduced,
only $v_{41}$ and $v_{43}$ are different from zero,  \bea v_{41} =
v_{43} =  -\frac{\Omega_d \Lambda^{d-3}}{48 (2\pi)^d} \;\lambda \;
\delta(t_1-t_2) \delta(t_1-t_3) \delta(t_1-t_4).\ena  (The factor
$\Lambda^{d-3}$, actually equal to 1, has been included for
completeness.)}

According to the definitions (\ref{S2}) and (\ref{Sn}), $v_{nm}$
couples $n$ fields altogether, $m$ of type $\sqp$ and $n-m$ of
type $\cp$, always with $m\ge 1$. In addition, when in a given
term the number $m$ of fields of the type $\sqp$ is even, the
coupling function appears multiplied by $\mi$. Therefore all the
couplings functions $v_{nm}$ will be real [see Eq. (\ref{propiedad
2})]. We will take the couplings symmetrical with respect to the
permutations of the variables of each type of field; e.g.,
$v_{42}$ is unchanged when $(\bfk_1, t_1) \leftrightarrow (\bfk_2,
t_2)$ or $(\bfk_3, t_3) \leftrightarrow (\bfk_4, t_4)$. We will
also assume that the $v_{nm}$ couplings do not contain time
derivatives of Dirac deltas.

To make sure that the RG transformation relates actions on the
same class and preserves the CTP properties (\ref{propiedad 1})
and (\ref{propiedad 2}), the couplings $v_{nm}$ must be such that
at least one of the $\sqp$ fields in each interaction term is
evaluated at a time $t$ equal or later than  the fields $\cp$.
This condition  is directly related to the fact that the
expectation value $\lt<\cp(t) \sqp(t')\rt>$ is causal.

Finally, we give a prescription to separate out the free action
from the quadratic interaction terms. It must be
\bea &\D\int_0^T dt \D\int_0^T dt'\; v_{21}(0; t,t') = 0, &\nn \\ \nn \\
\nn &\D\frac{\partial^2}{\partial k^2}\lt(\D\int_0^T dt \D\int_0^T
 dt'\; v_{21}(k; t,t')\rt)\bigg|_{k=0} = 0, & \\ \nn \\ &\D\int_0^T
dt \D\int_0^T   dt'\; (t-t')\; v_{21}(k; t,t') = 0,& \nn  \\ \nn \\
&\D\int_0^T dt \D\int_0^T  dt'\; v_{22}(k; t,t') = 0.& \ena In
this way, the terms $m^2 \sqp \cp$, $k^2 \sqp \cp$, $\kappa \sqp
\dot \cp$ and $\nu \sqp^2$ are isolated within $S_0$. We will
return to this point in Appendix \ref{obtaining delta S}.

We have to chose the initial conditions $\rho[\sqp(\bfk,0),
\cp(\bfk, 0)]$. We will use \bea &\rho[\sqp(\bfk,0), \cp(\bfk, 0)]
= \exp\Bigg\{-\D\int_\Lambda d^d\!k \;\frac{a(k)}{4} \Bigg[
\tanh\lt(\frac{a(k)
\beta(k)}{2}\rt)\; \sqp(\bfk, 0) \sqp(-\bfk, 0) & \nn \\ \nn \\
 &+\coth\lt(\D\frac{a(k) \beta(k)}{2}\rt) \;\cp(\bfk, 0)
\cp(-\bfk, 0) \Bigg] \Bigg\}. & \label{initial density matrix}
\ena {This corresponds to uncoupled, free fields with frequency
$a(k)$ and with a temperature $\beta(k)^{-1}$ which depends on the
wave number. Observe, however, that this is a nonequilibrium
initial condition for the interacting fields.}

The propagators corresponding to the free action (\ref{the action
2}) and the density matrix (\ref{initial density matrix}) are
given in Appendix \ref{appendix propagators}.

\section{The RG transformation \label{section the RG
transformation}}

{In this section, we define the RG transformation for the general
class of actions (\ref{the action 3}). The infinitesimal RG
transformation, from which the finite transformation is obtained,
is composed by two operations: i) elimination of the modes with
wave numbers in an infinitesimal momentum shell, and ii)
rescaling. We analyze each of these operations in the following
two subsections. The joint effect of i) and ii) is written as a
set of differential equations in Sec. \ref{section RG equations
and formal solutions}.}

\subsection{Mode elimination}

The first step to obtain the RG equations is discussed in Appendix
\ref{section scalar field considered}. Here we will apply the
results of that section to the action (\ref{the action 3}). Modes
with momenta in the shell between $b$ and $1$ are eliminated ($b =
1 - \delta s$, $0<\delta s \ll 1$). After eliminating the modes in
the shell, one obtains an effective action for the modes with
momenta below $b$. The effective action is given by Eqs.
(\ref{effective action}) and (\ref{delta S}). It can be written as
\bea S_< = S_0' + S_{\rm int}' \ena where \bea &\nn  S_0'[\sqp,
\cp] =\D\int_{0}^{T} dt \,\D\int_{b\Lambda} d^d\!k \;\;
\Bigg\{\D\frac{1}{2} \, \dot\sqp \, \dot \cp  - \D\frac12
\lt[b^{-2\eta} \;k^2 + (m^2+\delta m^2) \rt] \sqp\, \cp & \nn \\
\nn \\  & -(\kappa + \delta \kappa) \; \sqp \,\dot\cp +
\D\frac{i}{2} (\nu + \delta \nu)\; \sqp \sqp \Bigg\},& \label{S0
prima}\ena  and where $S_{\rm int}'$ has the same form that
$S_{\rm int}$ in Eq. (\ref{expansion Sint}), but with perturbed
couplings \bea \label{vnm prima} v_{nm}' = v_{nm} + \delta v_{nm}.
&\ena Note that the momenta in the integrals now go up to $b
\Lambda$. We discuss in the Appendix \ref{obtaining delta S} the
actual calculations involved.

At this stage, the initial density matrix remains unchanged.

\subsection{Rescaling \label{section rescaling}}

Now we introduce the second step in the RG transformation. We
redefine the fields and change integration variables in the
action, writing \bea \label{resc 1} b^{\alpha_\sqp} \sqp\lt(b^{-1}
\bfk, b^{\alpha_t} t\rt)\ena and \bea \label{resc 2}
b^{\alpha_\sqp} \cp\lt(b^{-1} \bfk, b^{\alpha_t} t\rt)\ena instead
of $\sqp(\bfk, t)$ and $\cp(\bfk, t)$, and \bea \label{resc 3} b
\bfk\ena and \bea b^{-\alpha_t} t \label{resc 4}\ena instead of
$\bfk$ and $t$. The rescaling redefines fields, times and momenta
and  restores the cutoff to its original value $\Lambda = 1$. The
two exponents, \bea \label{exponente campo} \alpha_\sqp &=&
\frac12 (\eta  - d - 1), \\ \label{exponente tiempo} \alpha_t &=&
1- \eta, \ena are chosen to reduce to 1 the coefficients of the
terms $\dot \sqp \dot \cp$ and $k^2 \sqp \cp$ in the free action.
Here $\eta$ is the quantity introduced in (\ref{S0 prima}).
Rescaling affects both the couplings in the action and the
parameters which define the density matrix.

{It is important to notice that, in spite of rescaling, the time
$T$ may be kept as a RG invariant. This is a consequence of the
CTP condition (\ref{CTP condition 2}) and causality. This fact and
other details concerning rescaling are shown in Appendix
\ref{appendix rescaling}.}

\subsection{RG equations and formal solutions
\label{section RG equations and formal solutions}}

The effect of mode elimination and rescaling over the parameters
which define the action and the density matrix can be summarized
in the following differential equations \bea \label{de m2}
\lt(\D\frac{\partial}{\partial s} -2+2\eta\rt)m^2 = \D\frac{\delta
m^2}{\delta s}, \ena \bea \label{de kappa}
\lt(\D\frac{\partial}{\partial s} - 1+\eta + k
\D\frac{\partial}{\partial k}\rt)\kappa = \D\frac{\delta
\kappa}{\delta s}, \ena \bea \label{de nu}
\lt(\D\frac{\partial}{\partial s}-2+2\eta +
k\D\frac{\partial}{\partial k}\rt)\nu = \D\frac{\delta \nu}{\delta
s}, \ena  \bea \lt[\D\frac{\partial}{\partial s} -
\lt(\D\frac{n}{2}-1\rt) \, (3-d) + \D\frac{3n}{2} \eta -3 +
\D\sum_{i=1}^n \lt(\bfk_i \D\frac{\partial}{\partial \bfk_i}
-\alpha_t \; t_i \D\frac{\partial}{\partial t_i}\rt) \rt] v_{nm} =
\D\frac{\delta v_{nm}}{\delta s}, \label{de vnm} \ena \bea
\lt(\D\frac{\partial}{\partial s}-1 + \eta  + k
\D\frac{\partial}{\partial k} \rt) a &=& 0, \label{de a}\\ \nn \\
\lt(\D\frac{\partial}{\partial s}+ 1-\eta  + k
\D\frac{\partial}{\partial k} \rt) \beta &=& 0. \label{de
beta}\ena [See Eqs. (\ref{ec diferencias m2})-(\ref{ec diferencias
beta}). Equations. (\ref{exponente campo}) and (\ref{exponente
tiempo}) have been used for the exponents.]

Equations (\ref{de m2})-(\ref{de beta}), together with the
equation for $\eta$, are the RG equations. Integrating these
differential equations is equivalent to iterate the infinitesimal
transformation. They have the general form \bea \label{general
form} \lt[\D\frac{\partial}{\partial s} + \alpha_F +
\D\sum_{i=1}^n \lt(k_i \D\frac{\partial}{\partial k_i} -\alpha_t
\; t_i \D\frac{\partial}{\partial t_i}\rt)\rt] F = g, \ena where
$\alpha_F$ and $\alpha_t$ are functions of $s$ alone. The formal
solution $F(\{\bfk\}; \{t\}, s)$,  with the initial condition \bea
F\lt(\{\bfk\}; \{t\}; 0\rt) = F_0(\{\bfk\}; \{t\}), \ena is \bea
\nn & F(\{\bfk\}; \{t\}, s) = \D\int_0^s ds' \;
e^{\beta_F(s')-\beta_F(s)} \; g\lt(e^{s'-s}
\{\bfk\}; e^{\beta_t(s)-\beta_t(s')} \{t\}, s'\rt) + & \nn  \\
& e^{-\beta_F(s)}  F_0\lt(e^{-s} \{\bfk\}; e^{\beta_t(s)}
\{t\}\rt),&  \label{formal solution} \ena where $\beta_F(s) =
\D\int_0^s ds' \; \alpha_F(s')$, and  $\beta_t(s) = \D\int_0^s ds'
\; \alpha_t(s')$.

\section{The RG equations to order $\lambda^2$
\label{section the RG equations to order lambda2}}

\subsection{Reduced set of parameters}

{So far, the RG transformation defined for the class of actions
(\ref{the action 3}) is rather general. The transformation is
closed with respect to this class, but infinitely many couplings
have to be taken into account. To reduce the number of couplings
that have to be considered, we will choose as the initial
condition at the cutoff $\Lambda$ the $\lambda \phi^4$ action
(\ref{the action 1}), and compute the RG equations to order
$\lambda^2$. The central question is to find the minimum set of
parameters such that the transformation is closed to order
$\lambda^2$ for this initial condition. This set, which has 18
elements, is found in Appendix \ref{appendix V}, together with a
number of constraints upon the dependencies on the momenta and
time variables.}

These results can be summarized by saying that, starting at $s =
0$ from the action (\ref{the action 1}), to order $\lambda^2$, the
action for $s > 0$ can be written as $S = S_0 + S_{\rm int}$,
where $S_0$ is the free action of Eq. (\ref{the action 2}), and
where $S_{\rm int}$ is given by  \bea \nn & S_{\rm int} =
\D\int_{0}^{T} dt \, \D\int_\Lambda d^d\!k \; V_{21}(t,s) \;\,
\sqp(\bfk, t) \cp(-\bfk, t)  & \\ \nn &+ \D\int_{0}^{T} dt
\D\int_0^T  dt' \, \D\int_\Lambda d^d\!k \; \Big[W_{21}(k; t,
t',s) \;\, \sqp(\bfk, t) \cp(-\bfk, t') + i W_{22}(k; t, t',s)
\;\, \sqp(\bfk, t) \sqp(-\bfk, t') \Big]  & \nn
\\  \nn & + \Omega_d^{-1} \D\int_{0}^{T} \!\!dt \D\int_\Lambda
\D\prod_{i=1}^4 d^d\!k_i  \; \delta^d\lt(\bfk_1+\dots+\bfk_4\rt)
\Big[V_{41}(s) \;\, \sqp_1 \cp_2 \cp_3 \cp_4 + V_{43}(s)\;\,
\sqp_1 \sqp_2 \sqp_3 \cp_4 \Big]  &  \\ \nn &+ \Omega_d^{-1}
\D\int_{0}^{T} \!\!dt \D\int_0^T  \! dt' \D\int_\Lambda
\D\prod_{i=1}^4 d^d\!k_i  \; \delta^d\lt(\bfk_1+\dots+\bfk_4\rt)
\Big[W_{41}(Q_{12}; t, t',s) \;\, \sqp_1 \cp_2 \;\, \cp_3' \cp_4'
& \\ \nn & +i W_{42}(Q_{12}; t, t',s) \;\, \sqp_1 \cp_2 \;\,
\sqp_3' \cp_4' + W_{43}(Q_{12}; t, t', s) \;\, \sqp_1 \cp_2 \;\,
\sqp_3' \sqp_4' \Big]  & \\  \nn &+ \Omega_d^{-2} \D\int_{0}^{T}
\!\!dt \D\int_0^T \! dt' \D\int_\Lambda \D\prod_{i=1}^6 d^d\!k_i
\; \delta^d\lt(\bfk_1+\dots+\bfk_6\rt) \Big[v_{61}(Q_{123}; t,
t',s) \;\, \sqp_1 \cp_2 \cp_3 \;\, \cp_4'
 \cp_5' \cp_6'  & \\ \nn & + i v_{62}(Q_{123}; t, t',s) \;\, \sqp_1 \cp_2
\cp_3 \;\, \sqp_4' \cp_5' \cp_6' + v^{(1)}_{63}(Q_{123}; t, t',s)
\;\, \sqp_1 \sqp_2 \sqp_3 \;\, \cp_4' \cp_5'  \cp_6'  & \\ \nn  &
+ v^{(2)}_{63}(Q_{123}; t, t',s) \;\, \sqp_1 \cp_2 \cp_3 \;\,
\sqp_4' \sqp_5' \cp_6' + i v_{64}(Q_{123}; t, t',s) \;\, \sqp_1
\cp_2 \cp_3 \;\, \sqp_4' \sqp_5' \sqp_6'   & \\   & +
v_{65}(Q_{123}; t, t',s) \;\, \sqp_1 \sqp_2 \sqp_3 \;\, \sqp_4'
\sqp_5' \cp_6' +  i v_{66}(Q_{123}; t, t',s) \;\, \sqp_1 \sqp_2
\sqp_3 \;\, \sqp_4' \sqp_5' \sqp_6' \Big].& \ena Here $\sqp_i =
\sqp(\bfk_i, t)$, $\sqp_i' = \sqp(\bfk_i, t')$ (same for $\cp$),
and \bea Q_{12\dots} = |\bfk_1+\bfk_2+\dots|.\ena Moreover, $m^2$,
$V_{21}$, $V_{41}$ and $V_{43}$ are of order $\lambda$, while
$\kappa$, $\nu$, the rest of the couplings in $S_{\rm int}$, and
$\eta$ are of order $\lambda^2$.

It is straightforward to recast the general analysis of Sec.
\ref{section the RG transformation} for this particular action.
The RG group equations are given in the next section.

\subsection{The RG equations}

{In the previous subsection we gave the reduced set of parameters
that has to be considered in working out the equations to order
$\lambda^2$. Now the RG equations will be grouped in the same
order that will be solved later on. The groups form a hierarchy.
The equations at the top are themselves a closed system. Once
their solutions are known, the equations at the following level
can be considered as effectively closed, and so on.

At the top of the hierarchy are the equations that determine the
propagators $\cG$ and $\mG$. It is important to notice that in
writing the RG equations to order $\lambda^2$, for most cases it
is enough to know the propagators $\cG$ and $\mG$ to order zero in
$\lambda$. At that order they are independent of the couplings and
are determined by the functions $a$ and $\beta$ in Eq.
(\ref{initial density matrix}) alone. The only exceptions are the
equations for $V_{21}$ and $m^2$, where $\cG$ must be known to
order $\lambda$, in which case it will depend on $m^2$. But these
equations are at the end of the hierarchy, so they do not affect
the equations that precede them and still will form a closed
system.

Thus, for the moment, we only need the propagators to order zero
in $\lambda$ evaluated at $k = \Lambda = 1$. From their general
expressions (\ref{prop mG}) and (\ref{prop cG}) we find \bea
\label{propagador 100} \mG(t, t') &=& 2 \sin(t-t') \;
\Theta(t-t'), \ena and \bea \label{propagador 200} &\cG(t,t',s) =
\D\frac{2}{a(1, s)}  \!\lt[1+2f\Big(a\,\beta(1, s)\Big)\rt] \!
\lt\{a(1, s)^2 \cos(t-t') + \lt[1-a(1, s)^2\rt]
\cos(t+t')\rt\}.&\ena  The functions $a(1,s)$ and $\beta(1,s)$ to
order zero in $\lambda$ are given by Eqs. (\ref{de a}) and
(\ref{de beta}) setting $\eta = 0$.}

At the second level are the equations for $V_{41}$ and $V_{43}$
\bea \label{de V41} \lt(\D\frac{\partial}{\partial s} + d - 3 \rt)
V_{41}(s)
&=& 18 V_{41}(s)^2 \, \Gg(s, T), \\
\nn \\ \label{de V43}  \lt(\D\frac{\partial}{\partial s} + d -
3\rt) V_{43}(s) &=& 18 V_{41} V_{43}(s) \, \Gg(s, T), \ena where
\bea \label{def Gg} \Gg = \cP (\mG\cG). \ena  It has been used
that $v_{6i}(1; t, t',s) = 0$ for the chosen initial conditions,
and that $\eta$ can be set equal to zero in the left-hand side
member of the RG equations to order $\lambda^2$ [see Eqs. (\ref{de
m2})-(\ref{de vnm})]. {Note that if $V_{41}(s) = V_{43}(s)$ for $s
= 0$, then this relation holds for $s > 0$ as well. The solutions
$V_{41}$ and $V_{43}$ will depend on $T$ through the function
$\Gg$.}

At the next level we find the equations for the $v_{6i}$
couplings,  Appendix \ref{appendix V}. They have the same general
form; we write only the first three equations, that are all that
is needed for present discussion \bea \label{de v61} \lt(\bfD + 2d
- 5 \rt) v_{61}(k; t, t',s) =3 V_{41}(s)^2 \, \mG(t,t') \,
\delta(k - 1^+), \ena \bea \label{de v62} \lt(\bfD + 2d - 5\rt)
v_{62}(k; t, t',s) = \D\frac92 V_{41}(s)^2 \, \cG(t,t',s) \,
\delta(k - 1^+), \ena \bea \label{de v631} \lt(\bfD + 2d - 5\rt)
v^{(1)}_{63}(k; t, t',s) = 9 V_{41}(s) V_{43}(s) \, \mG(t,t') \,
\delta(k - 1^+). \ena [See Eq. (\ref{def delta +}) for a
definition of $\delta(k-1^+)$.] Here \bea \bfD &=&
\D\frac{\partial}{\partial s} + k \D\frac{\partial}{\partial k} -
\lt(t\D\frac{\partial}{\partial t} + t'\D\frac{\partial}{\partial
t'} \rt). \ena According to the results of Appendix \ref{appendix
V}, the factor $\alpha_t$, which would have to appear in front of
the time derivatives, has been set equal to one.

Next come the equations for $W_{41}$, $W_{42}$ and $W_{43}$, which
are \bea & \lt(\bfD + d-4 \rt) W_{41}(k; t, t',s) = 18
\delta_{k;0} \; V_{41}(s)^2 \; \mG\cG(t,t',s) - 18 V_{41}^2 \, \Gg
\; \delta(t-t') & \nn \\  \label{de W41} &+ \D\int
\D\frac{d\Omega_d}{\Omega_d} \lt[6 v_{61}(Q_{k\Omega}; t, t',s) \,
\cG(t,t',s)  +  2 v_{62}(Q_{k\Omega}; t, t', s) \, \mG(t,t')\rt],&
\ena \bea &\lt(\bfD + d -4 \rt) W_{42}(k; t, t',s) = 9
\delta_{k;0} \; V_{41}(s) \Big[V_{41}(s) \, \cG\cG(t,t',s) \, - 2
V_{43}(s) \lt[\mG\mG(t,t')\rt]_{\overline {tt'}} \Big] & \nn \\
\label{de W42} &-\D\int \D\frac{d\Omega_d}{\Omega_d} \lt\{4
\lt[v^{(1)}_{63}(Q_{k\Omega};t,t',s)\, \mG(t,t')
\rt]_{\overline{tt'}} +  4 v_{62}(Q_{k\Omega}; t, t',s) \,
\cG(t,t',s)\rt\}, & \ena  \bea & \lt(\bfD + d - 4 \rt) W_{43}(k;
t, t',s) = 18  \delta_{k;0}  \; V_{41} V_{43}(s) \; \mG\cG(t,t',s)
- 18 V_{41} V_{43}(s) \, \Gg(s) \; \delta(t-t')  &  \nn \\
\label{de W43} & +\D\int \D\frac{d\Omega_d}{\Omega_d}\lt[2
v^{(1)}_{63}(Q_{k\Omega}; t, t',s) \, \cG(t,t',s) + 6
v_{64}(Q_{k\Omega}; t, t',s) \, \mG(t,t')\rt].& \ena We defined
\bea Q_{k\Omega} = |\bfk + \hat \Omega|. \ena The subscript
$\overline {tt'}$ means symmetrization respect to $t$ and $t'$ (we
used that $v_{62}$ and $\cG$ are already symmetric in these
variables). The  terms in Eqs. (\ref{de W41}) and (\ref{de W43})
proportional to $\delta(t-t')$ are the necessary  subtractions
after isolating the contributions to $\delta V_{41}$ and $\delta
V_{43}$.

Finally, we write the equations for the the 2 field couplings and
$\eta$, which can be solved in closed form if the solutions of the
previous equations are known.  We define three auxiliary functions
\bea \label{def f1} f_1(t,s) = \int_0^t dt' \Big[W_{41}(0; t,
t',s) \, \cG(t',t',s) + V_{41}(s) V_{21}(t',s) \, \mG\cG(t,t',s)
\Big] + 3 V_{41}(s) \, \cG(t,t,s), \ena \bea \label{def f2} f_2(k;
t, t',s) = 2 \D\int\D\frac{d\Omega_d}{\Omega_d} \Big[
W_{41}(Q_{k\Omega}; t, t',s) \, \cG(t,t',s) + W_{42}(Q_{k\Omega};
t, t',s) \, \mG(t,t') \Big], \ena \bea \label{def f3} f_3(k; t,
t',s) = \D\int\D\frac{d\Omega_d}{\Omega_d} \lt[W_{42}(Q_{k\Omega};
t,t',s) \, \cG(t,t',s) + 2 W_{43}(Q_{k\Omega}; t, t',s) \,
\mG(t,t')\rt]. \ena Each function is the sum of several diagrams
appearing in Appendix \ref{appendix V}.

In terms of these functions, it results \bea \label{de V21}
\lt(\bfD - 2 \rt) V_{21}(t,s) = f_1(t,s) - \cP f_1(s), \ena \bea &
\lt(\bfD - 3\rt) W_{21}(k; t, t',s) = f_2(k; t, t',s) - \lt[\cP
f_2(0,s) + \D\frac{1}{2} k^2 (\cP f_2)''(0,s)\rt] \delta(t-t')  &
\nn \\ \nn \\ \label{de W21} & -\cQ f_2(k,s) \; \lt\{2
\lt[\D\frac{\partial }{\partial t'} + \delta(t') - \delta(0)\rt]
\delta(t-t')\rt\},&\ena \bea \label{de W22} \lt(\bfD - 3 \rt)
W_{22}(k; t, t',s) = f_3(k; t, t',s) - \cP f_3(k,s) \;
\delta(t-t'), \ena  \bea \label{de m2 2}\lt(\bfD - 2 \rt) m^2(s) =
-2 \lt[\cP f_1(s) + \cP f_2(0,s) \rt], \ena \bea \label{de kappa
2} \lt(\bfD - 1 \rt) \kappa(k,s) = - \cQ f_2(k,s), \ena \bea
\label{de nu 2}\lt(\bfD - 2 \rt) \nu(k,s) = 2 \, \cP f_3(k,s),
\ena \bea \label{eta 2} \eta = -\D\frac{1}{2} \,
\D\frac{\partial^2 \cP f_2}{\partial k^2}(0,s). \ena The operators
$\cP$ and $\cQ$ are defined in Appendix \ref{obtaining delta S},
Eqs. (\ref{def cP}) and (\ref{def cQ}). {Note also that, as was
mentioned at the beginning of this subsection,  since $V_{41}$ is
of order $\lambda$, in the last term of Eq. (\ref{def f1}), $\cG$
must be written to order $\lambda$. In this way, the equations for
$V_{21}$ and $m^2$ will include all the terms to order
$\lambda^2$. As we will not attempt to find the solutions for
$V_{21}$ and $m^2$ to order $\lambda^2$ we will not need to write
$\cG$ to order $\lambda$ explicitly.}

The way to integrate the RG equations starts by writing down the
propagators $\mG$ and $\cG$ as functions of $s$. Next the
equations for $V_{41}$ and $V_{43}$ can be integrated immediately.
Once these functions are known, the equations for the $v_{6i}$
couplings are solved using the formal solution (\ref{formal
solution}). With these solutions, the equations for the $W_{4i}$
can be integrated, and then the equations for  $V_{21}$, $W_{2i}$,
$m^2$, $\kappa$, $\nu$ and $\eta$. This is immediate, except for
$V_{21}$ and $m^2$, because they are given by integro-differential
equations. (They become ordinary differential equations to order
$\lambda$.)

We will work in $d = 3$. First we find the propagators to order
zero in $\lambda$. We need  initial conditions for $a$ and $\beta$
in the density matrix. We choose \bea \label{a inicial} a(k, 0) =
\omega_0(k, 0) = k,\ena and \bea \label{beta inicial} \beta(k, 0)
= \beta_0(k). \ena For the moment $\beta_0$ can remain
unspecified. From Eq. (\ref{de a}), with $\eta = 0$ \bea a(k, s) =
k .\ena Note that the evolution  of the product $a \beta$, that
appears in the argument of the function $f$ in Eq.
(\ref{propagador 2}) for $\cG$, is given by \bea
\lt(\D\frac{\partial}{\partial s} + k \D\frac{\partial}{\partial
k} \rt) a \beta &=& 0 \ena which has the immediate solution \bea
a\beta(k, s) = a\beta(k e^{-s}, 0) \label{sol abeta}. \ena It is
convenient to use the variable $z = e^s$ instead of $s$ in the
propagators. Using (\ref{a inicial}) and (\ref{beta inicial}) in
(\ref{sol abeta}), and then replacing in (\ref{propagador 200}),
setting  $k = 1$ \bea \label{propagador 21} \cG(t,t',z) = 2 \lt[1
+ 2 f\Big(z^{-1} \beta_0(z^{-1})\Big) \rt] \cos(t-t'), \ena from
which Eq. (\ref{def Gg}) gives \bea \label{Gg 2} \Gg(z, T) = \lt[1
- \frac1{2T} \, \sin(2T)\rt] \; \lt[1 + 2 f\Big(z^{-1}
\beta_0(z^{-1})\Big) \rt]. \ena The dependence of $\Gg$ on $T$
comes through the first term in Eq. (\ref{Gg 2}). In Fig. 1 we
plot $1 - (2T)^{-1} \, \sin(2T)$ as a function of $x = 1/T$.
Observe that for $T \ra \infty$ the function tends to 1.

\begin{figure}
\includegraphics[angle=-90, width = 6cm]{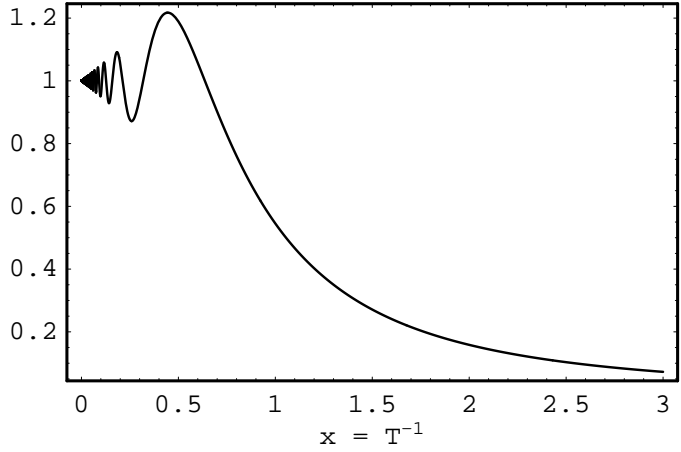}
\label{graph Gg} \caption{{\footnotesize The graph of $1 -
(2T)^{-1} \, \sin(2T)$ as a function of $x = T^{-1}$. The term
$\Gg$ depends on $T$ through this function and has a definite
limit when $T  \ra \infty$.}}
\end{figure}

Then, from Eqs. (\ref{de V41}) and (\ref{de V43}), to order
$\lambda^2$, we get \bea V_{41}(s) = V_{43}(s) = V_0 + 18 V_0^2 \;
\int_0^s ds' \; \Gg(e^{s'}, T) ,\ena where \bea V_0 = -
\D\frac{\Omega_d}{(2\pi)^d} \; \D\frac{\lambda}{48}\bigg|_{d=3} =
-\D\frac{1}{2\pi^2} \; \D\frac{\lambda}{48}. \ena

To solve the equations for the $v_{6i}$ couplings to order
$\lambda^2$, in the second members of Eqs. (\ref{de v61})-(\ref{de
v631}) we can use the initial values of $V_{41}$ and $V_{43}$.
Using (\ref{formal solution}) \bea \label{sol v61} v_{61}(k; t,
t', s) = \frac{3}{k} \, V_0^2 \;
\mG\big(k (t, t')\big) \; \Theta(1^+ e^s - k) \Theta(k - 1^+), \\
\nn \\ \label{sol v62} v_{62}(k; t, t', s)= \frac{9}{2k} \, V_0^2
\; \cG\lt(k (t, t'),\D\frac{e^s}{k} \rt) \; \Theta(1^+ e^s - k)
\Theta(k - 1^+), \\ \nn \\ \label{sol v631} v^{(1)}_{63}(k; t, t',
s)= \frac{9}{k} \, V_0^2 \; \mG\big(k (t, t')\big) \; \Theta(1^+
e^s - k) \Theta(k - 1^+).\ena  [See Eq. (\ref{delta modulo}) for a
definition of $\Theta(k-1^+)$.] The notation $\Big(k (t, t')\Big)$
means $(k t, k t')$, and will be used frequently.

Next the equations for the $W_{4i}$ are integrated: \bea &
W_{41}(k; t, t',s) =  18\,\delta_{k; 0} \;  V_0^2 \;
\cI_1[\mG\cG](t,t', e^s)   &   \nn \\ \label{sol W41} & +9 V_0^2
\; \bigg[2 \, \cI_2[\cG, \mG](k; t, t', e^s)  + \cI_2[\mG, \cG](k,
t, t', e^s) \bigg] -  \Big[V_{41}(s) - V_0\Big] \; \delta(t-t'), &
\ena  \bea &W_{42}(k; t, t', s) =  9 \,\delta_{k; 0} \; V_0^2
\bigg[\cI_1[\cG\cG](t,t',e^s) -
2\cI_1[\mG\mG](t,t',e^s)_{\overline{tt'}}\bigg] & \nn \\
\label{sol W42}  & + 18 \, V_0^2 \; \bigg[\cI_2[\cG, \cG](k; t,
t',e^s)- 2\; \cI_2[\mG, \mG](k; t, t',e^s)_{\overline{tt'}}
\,\bigg], & \ena \bea &W_{43}(k; t, t',s) = 18 \, \delta_{k; 0} \;
V_0^2 \; \cI_1[\mG\cG](t,t',e^s) & \nn \\ \label{sol W43} & + 18
\, V_0^2 \bigg[\cI_2[\cG, \mG](k; t, t',e^s) + \cI_2[\mG, \cG](k;
t, t',e^s) \bigg] - \Big[V_{43}(s)-V_0\Big] \; \delta(t-t'), &
\ena where \bea \cI_1[G_1 G_2](t,t',z) = z\, \int_0^{s} \! ds' \;
e^{-s'} \; G_1G_2\bigg(z \, e^{-s'} (t,t'), e^{s'} \bigg), \ena
and \bea &\cI_2[G_1, G_2](k; t, t', z) = z \; \D\int_0^s ds'\;
e^{-s'} \; G_1 \bigg(z \,e^{-s'} (t,t'), e^{s'} \bigg)& \nn
\\ \nn \\ \label{definition of I2} &\times \D\int_\cS \D\frac{d\Omega_3}{\Omega_3} \, \D\frac{
G_2\lt(|z^{-1} \, e^{s'} \, \bfk + \hat \Omega| \, z e^{-s'}
(t,t'), |z^{-1} \, e^{s'} \, \bfk + \hat \Omega|^{-1} \,e^{s'}
\rt)}{|z^{-1} \,e^{s'} \, \bfk + \hat \Omega|}.& \ena Here, the
angular integration is over the region $\cS$ defined by \bea
{1<\lt|z^{-1} \, e^{s'}\, \bfk + \hat \Omega\rt| < e^{s'}}, \ena
so $\cI_2=0$ for $k=0$ (though its limit when $k\ra 0^+$ is not
necessarily 0). As before, $z$ is identified with $e^s$.

The function $\cI_2$ can be expanded as \bea \label{the function
I2 expanded} & \cI_2[G_1, G_2] = \Theta\lt(1+k-z\rt) \;
\cI_{2<}[G_1, G_2]+ \Theta(z-1-k)\; \cI_{2>}[G_1, G_2], \ena where
\bea \cI_{2<}[G_1, G_2] = \D\frac{1}{2k} \D\int_1^z dy \int_y^z dx
\; G_1 \bigg(y\, (t,t'),\D\frac{z}{y} \bigg) G_2\bigg(x  \;
(t,t'), \D\frac{z}{x} \bigg)\ena and \bea \cI_{2>}[G_1, G_2] =
\D\frac{1}{2k} \Bigg(\D\int_{1}^{z-k} dy \;\D\int_y^{y + k} dx +
\D\int_{z-k}^{z} dy \D\int_y^{z} dx \Bigg)  \; G_1 \bigg(y\,
(t,t'),\D\frac{z}{y} \bigg) G_2\bigg(x  \; (t,t'), \D\frac{z}{x}
\bigg). & \ena $\cI_2$ is continuous and have continuous first
partial derivatives along the line $z = k+1$.

We end this section taking  the limit of $T\ra\infty$, and zero
initial temperature, i.e. $\beta_0^{-1} = 0$. We focus in the
equations for $\lambda$ and $m^2$. In this limit, Eq. (\ref{Gg 2})
gives \bea \Gg = 1 \;\;\;\;\;\;\;\;\;\;(T\ra \infty, \;\;
\beta_0^{-1} = 0). \ena With initial conditions \bea V_{41}(0) =
V_{43}(0) = - \D\frac{\Omega_d}{(2\pi)^d} \; \D\frac{\lambda}{48},
\ena  the equations for $V_{41}$ and $V_{43}$ are  \bea \label{de
V41 limit} \lt(\D\frac{\partial}{\partial s} + d - 3 \rt)
V_{41}(s) &=& 18 V_{41}(s)^2, \ena \bea V_{43}(s) = V_{41}(s).
\ena For $s > 0$ we can define \bea \lambda(s) = - \frac{48
(2\pi)^d}{\Omega_d} V_{41}(s), \ena which will satisfies the
following differential equation \bea \lt(\D\frac{\partial
}{\partial s} + d - 3\rt) \lambda = -\D\frac{3\Omega_d}{8(2\pi)^d}
\lambda^2. \label{eq. lambda usual} \ena On the other hand, to
order $\lambda$, the equation (\ref{de m2 2}) for $m^2$  is \bea
\lt(\D\frac{\partial}{\partial s} - 2\rt) m^2 =
\D\frac{\Omega_d}{4 (2\pi)^d}  \, \lambda. \ena In computing $\cP
f_1$ to order $\lambda$ in Eq. (\ref{de m2 2}), it has been used
that \bea \cP \cG = 2 \;\;\;\;\;\;\;\;\;\;(T\ra \infty, \;\;
\beta_0^{-1} = 0). \label{eq. m2 usual} \ena Equations (\ref{eq.
lambda usual}) and (\ref{eq. m2 usual}) are the CTP equivalents of
the textbook equations derived using the RG group defined for an
Euclidean action \cite{Peskin}. Observe that we are integrating
out momentum shells of infinitesimal width in the spatial
directions but leaving the time direction unrestricted. Thus the
numerical coefficients in Eqs. (\ref{eq. lambda usual}) and
(\ref{eq. m2 usual}) may be different from those obtained when a
spherical shell in Euclidean momentum space is integrated out
\cite{Liao:Polonyi:Xu}.

\section{The damping constant $\kappa$ and the noise kernel $\nu$ \label{section kappa and
nu}}

\subsection{The damping constant}

To compute $\kappa$ we need the function $f_2$, Eq. (\ref{def
f2}), and then $\cQ f_2$. Not all the terms in $W_{41}$ and
$W_{42}$, given in Eqs. (\ref{sol W41}) and (\ref{sol W42}),
contribute to $\kappa$. The terms proportional to $\delta(t-t')$,
for which the application of $\cQ$ gives zero, can be discarded,
as well as the terms proportional to the Kronecker deltas, because
they do not contribute to the angular integrals. If Eq. (\ref{de
kappa 2}) is rewritten retaining only the relevant terms it yields
\bea (\bfD - 1) \kappa = -18 V_0^2 \, \cQ f_2^*, \ena where  \bea
& f_2^*(k; t, t', z) = \D\int\D\frac{d\Omega_3}{\Omega_3} \bigg[2
\, \cI_2[\cG, \mG](Q_{k\Omega}; t, t',z) + \cI_2[\mG,
\cG](Q_{k\Omega}; t, t',z) \bigg] \cG(t, t',z)
 &\nn
\\ \nn \\  & + \D\int\D\frac{d\Omega_3}{\Omega_3} 2 \bigg[\cI_2[\cG, \cG](Q_{k\Omega}; t, t',z)
- \cI_2[\mG, \mG](Q_{k\Omega}; t, t')\,\bigg] \mG(t,t').& \ena {In
the last line, it has been used that for the function $\mG$ given
in the Eq. (\ref{propagador 100}) is \bea
\cI_2[\mG,\mG](k;t,t',e^s)_{\overline{tt'}} \; \mG(t,t') =
\D\frac{1}{2} \, \cI_2[\mG, \mG](k; t, t',e^s) \; \mG(t,t'). \ena}
Using the formal solution in Eq. (\ref{formal solution}) \bea
\kappa(k, s) = -18 V_0^2 \, \int_0^s ds' \; e^{s-s'} \; \cQ
f_2^*(e^{s'-s} \, k, e^{s'}). \ena Due to the general structure of
the functions $\cI_2$, Eq. (\ref{the function I2 expanded}),
$\kappa$ will be given by an expression of the form \bea
&\kappa(k, s) = 18 V_0^2\, \D\int_0^s ds' \; e^{s-s'} \D\int
\D\frac{d\Omega_3}{\Omega_3} \bigg[\Theta\lt(|\,k\; e^{s'-s} +
\hat \Omega| + 1 - e^{s'}\rt)
A_{<}(|\,k \;e^{s'-s} + \hat \Omega|, e^{s'})  & \nn \\ \nn \\
\label{expresion kappa 1} & +\Theta\lt(e^{s'} - |\,k\; e^{s'-s} +
\hat \Omega| - 1\rt) A_>(|\,k \; e^{s'-s} + \hat \Omega|, e^{s'})
\bigg], & \ena with \bea A_< = -\cQ\bigg[\Big(2\cI_{2<}[\cG,\mG] \
+ \cI_{2<}[\mG,\cG]\Big) \cG + 2\Big(\cI_{2<}[\cG,\cG] -
\cI_{2<}[\mG,\mG]\Big) \mG \bigg], \ena and with a similar
expression for $A_>$. Observe that both $A_<$ and $A_>$ depend on
$T$ through the application of $\cQ$, and that this dependence
will be inherited by $\kappa$.

Changing variables, Eq. (\ref{expresion kappa 1}) can be rewritten
as \bea & \kappa(k, z) = 9 V_0^2 \, \D\frac{z^2}{k} \;
\D\int_{1}^{z} dy \D\int_{|1-\frac{y k}{z}|}^{1+\frac{y k}{z}} du
\; \D\frac{u}{y^3}\, \bigg[\Theta\lt(u+1-y\rt) A_<\lt(u, y\rt)  &
\nn \\ \nn
\\  & +\Theta\lt(y-u-1\rt) A_>\lt(u, y\rt) \bigg].&\ena In doing
this integral, there are three different regions of the plane $k
z$, restricted to  $0\le k \le 1$ and $1 \le z$, that must be
analyzed separately: i) $z \le 2 - k$, ii) $2-k<z\le 2+k$, and
iii) $2+k<z$. In each region $\kappa$ is given by the following
expressions:

\noindent Region i: \bea \kappa(k, z) = 9V_0^2\, \D\frac{z^2}{k}
\; \D\int_{1}^{z} dy \D\int_{1-\frac{y k}{z}}^{1+\frac{y k}{z}} du
\; \D\frac{u}{y^3} \;\, A_<\lt(u, y \rt) \label{zona i} .\ena

\noindent Region ii: \bea &\kappa(k, z) = 9V_0^2\, \D\frac{z^2}{k}
\; \Bigg( \D\int_{1}^{\frac{2z}{z+k}} dy \D\int_{1-\frac{y k}{z}
}^{1+\frac{y k}{z}} du  + \D\int_{\frac{2z}{z+k}}^z dy \D\int_{y
-1}^{1+\frac{y k}{z}}  du  \Bigg) \,  \D\frac{u}{y^3} \; A_<\lt(u,
y\rt)  &\nn \\ \nn \\ \label{zona ii} & +9V_0^2\, \D\frac{z^2}{k}
\,\D\int_{\frac{2z}{z+k}}^z dy \D\int_{1-\frac{y k}{z}}^{y-1} du
\;  \D\frac{u}{y^3} \; A_>\lt(u, y\rt). & \ena

\noindent Region iii: \bea &\kappa(k, z) =  9V_0^2\,
\D\frac{z^2}{k}\; \Bigg( \D\int_{1}^{\frac{2z}{z+k}} dy
\D\int_{1-\frac{y k}{z}}^{1+\frac{y k}{z}} du  +
\D\int_{\frac{2z}{z+k}}^{\frac{2z}{z-k}} \D\frac{dy}{y^3}
\D\int_{y-1}^{1+\frac{y k}{z}}  du  \Bigg) \,  \D\frac{u}{y^3} \;
A_<\lt(u, y\rt)  &\nn \\ \nn \\ \label{zona iii} &
+9V_0^2\,\D\frac{z^2}{k} \, \Bigg(
\D\int_{\frac{2z}{z+k}}^{\frac{2z}{z-k}} dy \D\int_{1-\frac{y
k}{z}}^{y-1} du +\D\int_{\frac{2z}{z-k}}^z dy \D\int_{1-\frac{y
k}{z}}^{1+\frac{y k}{z}} du \Bigg) \, \D\frac{u}{y^3} \; A_>\lt(u,
y\rt). & \ena

\noindent Note that for $k = 0$ the angular integral in Eq.
(\ref{expresion kappa 1}) is trivial \bea &\kappa(0, z) = 18 V_0^2
\, z \D\int_1^z \D\frac{dy}{y^2} \; A_<(1, y) \;\, \Theta(2 - z)
 & \nn \\ \nn \\ \label{kappa k cero} & +18V_0^2 \, z \lt[
\D\int_1^2 \D\frac{dy}{y^2} \; A_<(1, y) + \D\int_2^z
\D\frac{dy}{y^2} \; A_>(1, y) \rt] \;\, \Theta(z - 2), & \ena
which is also the limit when $k\ra 0$ of the expressions
(\ref{zona i}) and (\ref{zona iii}) for $z \le 2$ and $z > 2$,
respectively.

For $z \gg 1$ $\kappa$ becomes independent of $k$. If $A_>(1,y)$
grows slower than $y$, for $z \gg 1$ it results \bea \kappa(z)
\sim 18V_0^2 \, z \lt[ \D\int_1^2 \D\frac{dy}{y^2} \; A_<(1, y) +
\D\int_2^\infty \D\frac{dy}{y^2} \; A_>(1, y) \rt]= \kappa_0 \; z.
\ena This expression corresponds to what can be expected from
dimensional arguments. The constant $\kappa_0$ can be thought as
the $\kappa$ accumulated during the first stages of evolution. The
factor $z$ comes solely from the rescaling. If $A_>(1,y)$ grows as
$y$ or faster for $y\ra\infty$, to leading order, $\kappa$  will
be given by the last term in Eq. (\ref{kappa k cero}) \bea
\label{kappa z grande} \kappa(z) \sim 18 V_0^2 \; z \; \D\int_2^z
\D\frac{dy}{y^2} \; A_>(1, y). \ena Thus, in this case $\kappa$
will acquire an anomalous dimension.

\subsection{The noise kernel $\nu$}

The noise kernel $\nu$ is obtained following the same steps used
in the calculation of $\kappa$. First, Eq. (\ref{de nu 2}) is
rewritten keeping in $f_3$ only the relevant terms. Term
proportional to $\delta(t-t')$ in the solution (\ref{sol W43}) for
$W_{43}$ drops out  because $W_{43}$ appears multiplied by
$\mG(t,t')$, which is zero in $t = t'$). The terms proportional to
$\delta_{k;0}$ also drop out because they do not contribute to the
angular average. The equation for $\nu$ reduces to \bea \label{de
nu redux}\lt(\bfD - 2 \rt) \nu = 36 V_0^2 \, \cP f_3^*, \ena where
\bea & f_3^*(k; t, t', z) =  2
\D\int\D\frac{d\Omega_3}{\Omega_3}\bigg[\cI_2[\cG,
\mG](Q_{k\Omega}; t, t',z) + \cI_2[\mG, \cG](Q_{k\Omega}; t, t',z)
\bigg] \mG(t, t',z) &\nn \\ \nn \\  & +
\D\int\D\frac{d\Omega_3}{\Omega_3} \bigg[\cI_2[\cG,
\cG](Q_{k\Omega}; t, t',z) - \cI_2[\mG, \mG](Q_{k\Omega}; t,
t')\,\bigg] \cG(t,t').& \ena Defining  \bea &B_< =
\cP\bigg[2\Big(\cI_{2<}[\cG, \mG] + \cI_{2<}[\mG, \cG]\Big)  \mG +
\Big(\cI_{2<}[\cG, \cG] - \cI_{2<}[\mG, \mG]\Big)\cG\bigg],& \ena
and analogously $B_>$, it results  \bea & \nu(k, z) = 18 V_0^2 \,
\D\frac{z^3}{k} \; \, \D\int_{1}^{z} dy \D\int_{|1-\frac{y
k}{z}|}^{1+\frac{y k}{z}} du \; \D\frac{u}{y^4}\,
\bigg[\Theta\lt(u+1-y\rt) B_<\lt(u, y\rt)  & \nn \\ \nn \\  &
+\Theta\lt(y-u-1\rt) B_>\lt(u, y\rt) \bigg].&\ena Again, note that
the functions $B$ will depend on $T$ through the application of
$\cP$, an so will $\nu$.

Expressions analogous to Eqs. (\ref{zona i})-(\ref{zona iii}) hold
for $\nu$. For $k = 0$  \bea &\nu(0, z) = 36 V_0^2 \, z^2
\D\int_1^z \D\frac{dy}{y^3} \;
B_<(1, y) \;\, \Theta(2 - z)  & \nn \\ \nn \\
\label{nu k cero} & +36 V_0^2 \, z^2 \lt[ \D\int_1^2
\D\frac{dy}{y^3} \; B_<(1, y) + \D\int_2^z \D\frac{dy}{y^3} \;
B_>(1, y) \rt] \;\, \Theta(z - 2). & \ena For $z\gg 1$, $\nu$
becomes independent of $k$. If $B_>(1,y)$ grows slower than $y^2$,
the behavior of $\nu$ is controlled by the rescaling \bea \nu(z)
\sim 36 V_0^2 \, z^2 \lt[ \D\int_1^2 \D\frac{dy}{y^3} \; B_<(1, y)
+ \D\int_2^\infty \D\frac{dy}{y^3} \; B_>(1, y) \rt]. \ena
Otherwise $\nu$ acquires an anomalous dimension \bea \nu(z) \sim
36 V_0^2 \, z^2\, \D\int_2^z \D\frac{dy}{y^3} \; B_>(1, y)  \,. &
\ena

\subsection{$\kappa$ and $\nu$ for a specific choice of $\beta_0$}

We compute $\kappa$ and $\nu$ choosing $\beta_0^{-1}(k) = 0$,
namely the case of zero initial temperature. The function $1+2f$
in Eq. (\ref{propagador 21}), with $f$ defined in Eq. (\ref{efe}),
is replaced by 1. We compute $\kappa$ and $\nu$ only for $k = 0$,
since  the dependence on $k$ for $z \gg 1$ is weak. The detailed
expressions are given in Appendix \ref{appendix:detailed
expressions}.

The actual calculation shows that $A_>(1,y)$ and $B_>(1,y)$ both
grow as $y$ when $y\ra \infty$. This implies that, for $z \gg 1$,
$\kappa$ will develop an anomalous dimension, while $\nu$ will go
simply as $z^2$. Making explicit the dependence on $T$, for $z \gg
1$ \bea \nn \kappa(z, T) \sim  \D\frac{9V_0^2}{4T} \,\; z \log z
\;\; \lt[7 - 2T^2-8\cos T + \cos(2T)\rt]\ena and \begin{eqnarray}
\nonumber  \nu\left(k, T\right) \sim \frac{9 V_0^2}{8 T}\bigg\{\!
135+ 4 \Big[\gamma_{E}- 34\, \ln 2 - 7\, \ln 3+ 7 \,\Ci(3T)\Big]\! - 3 \, \cos(4T) \\
\nonumber  + \; 4\,\ln T   - 16\Big[8\pi + 3T - 16\,
\Si(2T)\Big] \sin T + 4\Big[7 \pi + 6 T - 14\, \Si(4T)\Big] \sin (2T) \\
\nonumber  -\, 12\Big[19 - 4T^2\Big]\, \Ci(T) + 8\Big[7-6T^2\Big]
\Ci(2T) + 4\Big[1-4T^2\Big]\Ci(4T) - 8\, \cos(3T)\\   - \;
24\big[\,7-8 \, \Ci(2T)\big] \cos T + 4\,\Big[11+2\, \Ci(T) - 2\,
\Ci(2T) - 14 \, \Ci(4T)\Big] \cos(2T) \nonumber  \\   + \;4T\big[
18\pi  - 54 \,\Si(T)+ 28\,\Si(2T) - 6 \,\Si(3T) - 2\, \Si(4T)
+\sin (4T)\big] \bigg\},\end{eqnarray} where $\Si$ and $\Ci$ are
the $\sin$ and the $\cos$ integral functions.  Observe that the
damping constant $\kappa$ is not definite positive.  This suggests
that the underlying mechanism could be similar to Landau damping
\cite{llpk}.

\section{Final Remarks}

In this paper we have investigated the nonequilibrium dynamics of
the low frequency, long wavelength modes of a self-interacting
real scalar field. We have computed the influence functional
encoding the back-reaction on these modes of the higher frequency,
shorter wavelength sector. We have obtained the coefficients of
the influence functional by solving the RG equations which
describe the change in the influence functional induced by the
progressive averaging over momentum shells.

The main finding of this paper is that the influence functional
for the low frequency modes contains terms associated with damping
and noise. It is most important to stress that we have not put
these terms by hand; contrariwise, we have assumed that they are
absent at high energies. Damping and noise are forced upon us by
the RG flow itself.

We have been able to retrieve these terms because we have gone
beyond the adiabatic approximation for the environment (high
frequency) modes. A crucial step in this direction is the
recognition of the role of the parameter $T$ in nonequilibrium
evolution. Because time-integration is restricted to the lapse
from $0$ to $T$, energy fluctuations, and thus particle creation
from the vacuum in the environment sector, are allowed even in
simple approximations to the dynamics. Although the finite
temperature RG in the real time formulation is well known, and
also the running of the effective potential in a CTP framework, to
the best of our knowledge this is the first instance were the RG
is used to compute an intrinsically nonequilibrium feature in
relativistic field theory.

The most limiting approximation we have made is the use of fixed
propagators in internal lines, disregarding the changes in the
propagators caused (mainly, but not only) by the generation of
mass, dissipation and noise terms through the RG flow. This
approximation forces us to restrict our analysis to very short
times (a few inverse cutoffs). A fully self-consistent RG should
overcome this shortcoming. Also we have only considered a very
simple -though nonequilibrium- initial condition for the field. A
more flexible approach regarding initial conditions is required
for most interesting applications.

The methods advanced in this paper are relevant to essentially all
the applications of nonequilibrium quantum field theory, from high
energy ones like thermalization in relativistic heavy ion
collisions and reheating after inflation to low energy
applications such as the dynamics of glasses and Bose-Einstein
condensates. It is clear that we have the bare outlines of a
framework, but we look forward to see this framework develop and
fructify in manifold ways.

\section*{Acknowledgments}

We acknowledge Joan Sola and  Enric Verdaguer for their kind
hospitality at Barcelona, and Enric Verdaguer for discussions.
This work was supported by Universidad de Buenos Aires, CONICET
and ANPCYT.

\appendix

\section{\label{section scalar field considered} Integrating out modes in a shell}

In the problem at hand,  the relevant system and the environment
are sectors of the same scalar field, being $\sqp_>$ and $\cp_>$
the variables associated with the short scales. The elimination of
these modes is the fundamental step in the RG transformation, and
follows the lines sketched in Sec. \ref{section open systems and
Feynman-Vernon formalism}.

The CTP generating functional is \bea Z[\sj, \cj] \!=\!\! \D\int
\!\!\cD\cp \cD \sqp \; \exp \mi \lt\{\! S_{\rm CTP}\lt[\sqp,
\cp\rt] \!+\!\! \D\int_{0}^{T} \!\!\! dt \D\int_\Lambda \!\!d^d\!k
\, \lt[\sj(\bfk, t) \cp(\bfk, t)  + \cj(\bfk, t) \sqp(\bfk, t)
\rt]\!\rt\} \rho[\sqp_0, \cp_0] \label{Z} . \ena  Here the
integral is over histories with $\sqp(\bfk, T) = 0$, and $\sqp_0$
and $\cp_0$ stand for the fields at $t = 0$. We retain explicitly
the dimensional parameter $T$ at which both histories on the CTP
integral coincide. The field configurations contain modes up to
momentum $\Lambda$: $\phi(\bfk, t)$ and $\cp(\bfk, t)$ are defined
in the region $|\bfk| = k \le \Lambda$, and so are the currents
$\sj$ and $\cj$. If necessary, fields and currents can be defined
to be zero outside this domain. The subscript in the momentum
integral indicates that the integration domain is bounded by $k =
\Lambda$.

The density matrix is known at $t = 0$. For simplicity we assume
that \bea \rho[\sqp, \cp] = \rho[\sqp_<, \cp_<] \otimes
\rho[\sqp_>, \cp_>],\ena with the same functional $\rho$
everywhere [cf. Eq. (\ref{initial density matrix}].

Only modes with wave numbers within an infinitesimal shell will be
integrated out. Let $0 < \delta s \ll 1$ the infinitesimal
parameter of the transformation, and define \bea b = 1 - \delta
s.\ena The momentum domain $k\le\Lambda$ is divided in two
regions: the shell \bea b \Lambda < k \le \Lambda,\ena and its
interior: \bea k\le b\Lambda.\ena Accordingly, the fields are
split in two parts, one which contains the modes within the shell,
indicated by $\phi_>$, and another part containing the modes with
$k \le b\Lambda$, indicated by $\phi_<$, i.e. \bea \nn \phi =
\phi_< + \phi_>, \ena where\bea \label{def phi menor} \phi_<(\bfk,
t) &=& \phi(\bfk, t)\, \Theta(b \Lambda  - k),
\\ \phi_>(\bfk, t) &=& \phi(\bfk, t)\, \lt[\Theta(k - b \Lambda)-\Theta(k-\Lambda)\rt]
\label{definicion phi mayor}, \ena and so for $\cp$. Here $\Theta$
is the unit step function. Equation (\ref{division}) reads \bea
\label{division 2} S_{\rm CTP}[\sqp, \cp] = S_{\rm CTP}[\sqp_<,
\cp_<] + \Delta S[\sqp_<, \cp_<, \sqp_>, \cp_>]. \ena  After
integrating out the modes within the shell, the effective action
for the surviving modes is \bea \label{effective action} S_{\rm
CTP<}[\sqp_<, \cp_<] = S_{\rm CTP}[\sqp_<, \cp_<] + \delta
S[\sqp_<, \cp_<],\ena where \bea \label{delta S} e^{i \delta
S[\sqp_<, \cp_<]} = \D\int \cD \cp_> \cD \sqp_> \; \exp\bigg(\mi
\Delta S[\sqp_<, \cp_<, \sqp_>, \cp_>]\bigg) \; \rho[\sqp_>(\bfk,
0), \cp_>(\bfk, 0)]. \ena This is essentially Eq. (\ref{SIF}). The
task here is to compute $\delta S$  to order $\delta s$.

\subsection{Computing $\delta S$ to order $\delta s$,
perturbative approach and Feynman rules \label{section
perturbative approach}}

We will compute $\delta S$ perturbatively. To do this, we will
separate in  $\Delta S$, Eq. (\ref{division 2}), a free CTP action
$S_0$, which will define the propagators to be used in the Feynman
diagrams, \bea \label{def SI} \Delta S[\sqp_<, \cp_<, \sqp_>,
\cp_>] = S_0[\sqp_>, \cp_>]+S_I[\sqp_<, \cp_<, \sqp_>, \cp_>].\ena
Momentarily and for sake of brevity, we will write $\sqp_>$
meaning the pair $\{\sqp_>$, $\cp_>\}$, and $\sqp_{>0}$ instead of
$\sqp_>(\bfk, 0)$ (analogous meaning for $\sqp_<$). Equation
(\ref{delta S}) reads \bea &e^{\mi \delta S[\phi_<]} = \D\int \cD
\phi_> \; \exp \mi \bigg(S_0[\phi_>]+S_I[\phi_<, \phi_>]\bigg) \;
\rho[\sqp_{>0}].& \ena This is identical to the usual expression
for a vacuum-vacuum amplitude. Hence,  \bea &\delta S[\phi_<] = -
i \log\lt(\D\int \cD \phi_> \; e^{\mi S_0[\phi_>]}\;
\rho[\sqp_{>0}] \rt) - \label{log}   i \D\sum_{n=1}^\infty
\D\frac{i^n}{n!} \lt<S_I^n\rt>_c[\phi_<]& \ena where

\vspace{-0.8cm}

\bea \label{promedios} \lt<S_I^n\rt>_c[\phi_<] = \D\frac{\D\int
\cD \phi_> \;  e^{\mi S_0[\phi_>]} \; \rho[\sqp_{>0}] \;
S_I[\phi_<, \phi_>]^n }{\D\int \cD \phi_> \; e^{\mi S_0[\phi_>]}
\; \rho[\sqp_{>0}]}\Bigg|_{\rm connected} \ena is the sum of the
connected diagrams drawn from $S_I^n$, with the fields $\sqp_<$
acting as external currents. The first, constant term in
(\ref{log}) can be omitted. Thus, essentially, we can write
(restoring the full notation for $\sqp$ and $\cp$) \bea
\label{delta S 2} \delta S[\sqp_<, \cp_<] =
\lt<S_I\rt>_c[\sqp_<,\cp_<] + \D\frac{i}{2!}
\lt<S_I^2\rt>_c[\sqp_<,\cp_<]+ \dots \ena  Internal lines in the
diagrams will correspond to the propagators deduced from
$S_0[\sqp, \cp]$. There are two propagators \bea \label{propagador
1} \lt<\cp(\bfk, t) \, \sqp(\bfk', t') \rt>_0 &=& -i \mG(k, t, t')
\, \delta^d(\bfk + \bfk'), \\  \label{propagador 2} \lt<\cp(\bfk,
t) \, \cp(\bfk', t') \rt>_0 &=& \cG(k, t, t')  \, \delta^d(\bfk +
\bfk').   \ena The subscript $0$ means that the expectation values
are taken with respect to the free action $S_0$. In general, the
function $\mG$ is zero if $t\le t'$. The third propagator
$\lt<\sqp \sqp'\rt>$ is always zero. These are two general
properties of the CTP formalism.

\begin{figure}
\includegraphics[width = 8 cm]{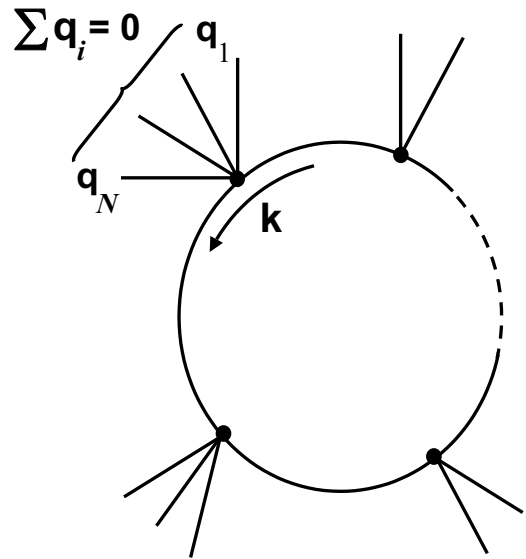}
\caption{{\footnotesize Type 1 diagrams have one loop with
momentum $k = \Lambda$. The total external momentum at each vertex
is zero.}} \label{type 1}
\end{figure}

\begin{figure}
\includegraphics[angle = -90, width = 10 cm]{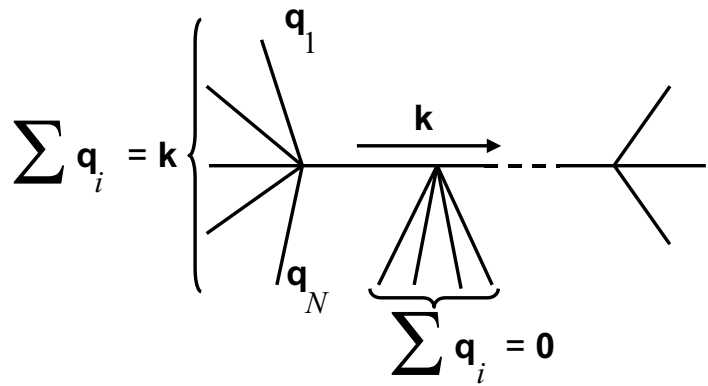}
\caption{{\footnotesize Type 2 diagrams are tree diagrams with no
external momentum entering at the intermediate vertices.}}
\label{type 2}
\end{figure}

It is clear that diagrams with more than one loop can be ignored
to order  $\delta s$. This is because for each loop there would be
one independent momentum integration over a region of volume of
order $\delta s$. Thus, the resulting term would be (at least)
proportional to $(\delta s)^L$, where $L$ is the number of loops
in the diagram.  Hence, it has to be $L \le 1$. In fact, diagrams
with $L = 0$ and $L = 1$ are both at least of order $\delta s$.
Owing to the fact that internal lines carry momenta in the
infinitesimal momentum shell, when computing $\delta S$ to order
$\delta s$, it is enough to consider two types of diagrams
\cite{WH}, shown in Fig. \ref{type 1} and Fig. \ref{type 2}.

\noindent Diagrams of the first type are shown in Fig. \ref{type
1}. They are one-loop diagrams. External lines attached to each
vertex must have total momentum equal to zero: for each vertex
with external lines carrying momenta $\bfq_1$, $\bfq_2$, $\dots$,
there will be a Kronecker delta \bea \delta_{Q; 0},\ena where \bea
\bfQ = \bfq_1 + \bfq_2 + \dots\ena Thus, the same momentum $\bfk =
\veco$, with $|\hat \Omega| = 1$, traverses the loop at every
point. The modulus of the momentum in the loop is fixed, but not
its direction $\hat \Omega$. Hence, instead of an volume
integration over $\bfk$, there will be just angular average over
$\hat \Omega$ and a multiplicative factor accounting for the
volume of the shell, that is \bea \Omega_d \Lambda^d \delta s \int
\D\frac{d\Omega_d}{\Omega_d}. \ena The fact that external lines at
each vertex must carry total momentum equal to zero can be
understood graphically. Consider for example the one-loop diagram
in Fig. 4.

\vspace{1.2cm}

\begin{center}
\begin{fmffile}{diag1}
\begin{fmfgraph*}(3.9,2.4)
   \fmfleftn{i}{2}
   \fmfrightn{o}{2}
   \fmf{plain,label=$$,label.side=left}{i1,v1}
   \fmf{plain,label=$$,label.side=right}{o1,v2}
   \fmf{plain,label=$$,label.side=right}{i2,v1}
   \fmf{plain,label=$$,label.side=left}{o2,v2}
   \fmf{plain,left,tag=1,label=$\bfk + \bfQ$,tension=1/2}{v1,v2}
   \fmf{plain,left,tag=1,label=$\bfk$,tension=1/2}{v2,v1}
   \fmfdot{v1,v2}

   \fmflabel{$\bfq_1$}{i2}
   \fmflabel{$\bfq_2$}{i1}
   \fmflabel{$\bfq_4$}{o1}
   \fmflabel{$\bfq_3$}{o2}

\end{fmfgraph*}
\end{fmffile}
\end{center}

\noi {\footnotesize FIG. 4. One-loop diagram used to illustrate
the fact that external lines should carry zero total momentum.
Here $\bfQ = \bfq_1 + \bfq_2$. Both $\bfk$ and $\bfk + \bfQ$ have
to be in the momentum shell.}

\vspace{5mm}

\noi The analytic expression for this diagram will include an
integration of the form \bea \int d^d\!k \; G_1(k)\, G_1(|\bfk +
\bfQ|) \; f(\bfq_1, \bfq_2, \bfq_3, \bfq_4, \bfk, -\bfk - \bfQ),
\label{volume} \ena where $\bfQ = \bfq_1 + \bfq_2$ is the external
momentum entering at the left vertex, and $G_1$ and $G_2$ are
propagators for modes of the field in the momentum shell $b\Lambda
< k \le \Lambda$. Then,  two conditions will determine the
integration volume, namely \bea b\Lambda < k \le \Lambda \ena and
\bea b\Lambda < |\bfk + \bfQ| \le \Lambda. \ena These conditions
are depicted in Fig. \ref{shells}. \setcounter{figure}{4}
\begin{figure}
\includegraphics[angle = -90, width = 6 cm]{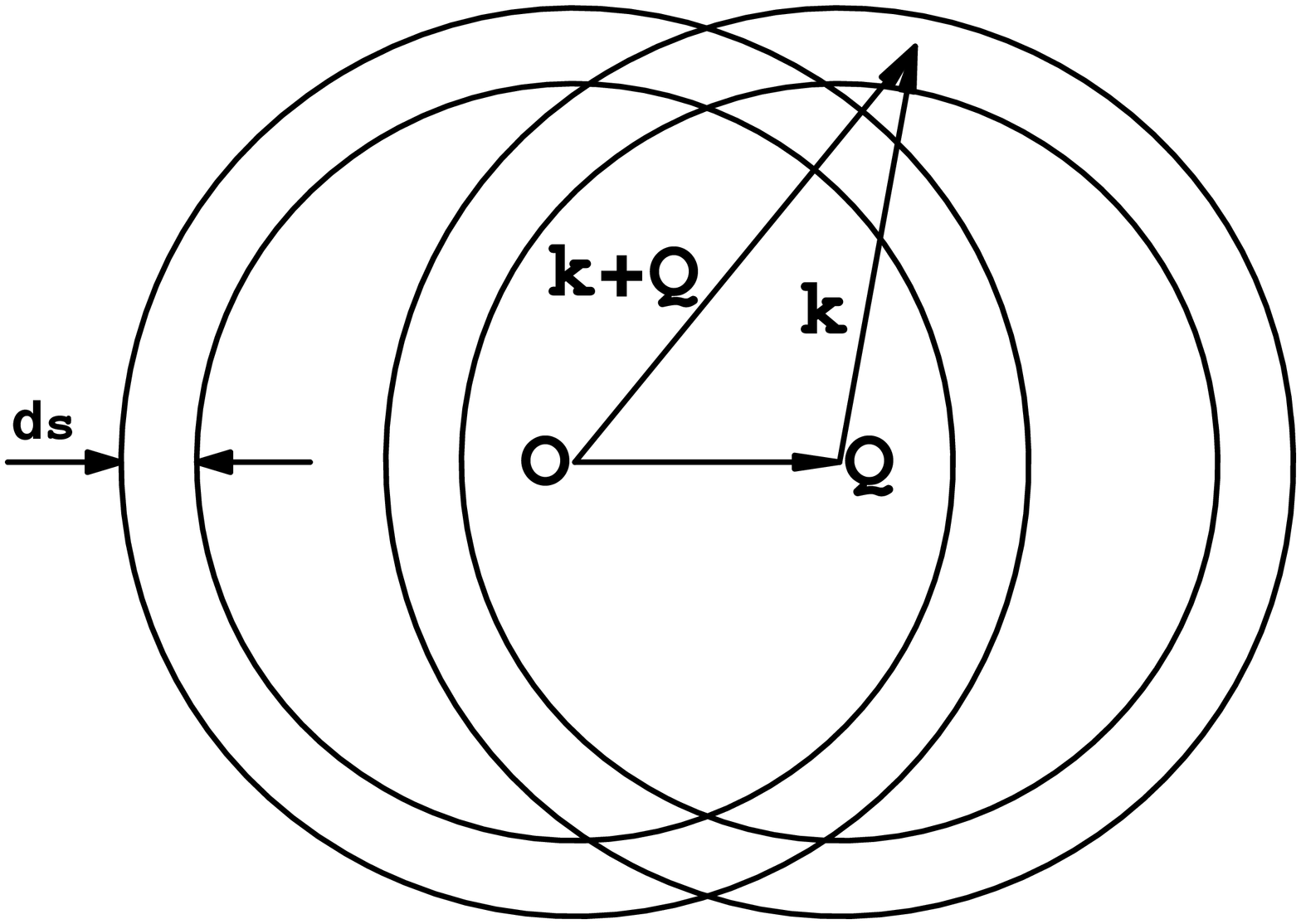}
\caption{{\footnotesize In $d = 2$, the integration region for the
diagram in Fig. 4    is the intersection of the two shells. Its
volume is of order $\delta s^2$ unless $Q \sim \delta s$.}}
\label{shells}
\end{figure}
\noi For $\delta s \ll 1$, the volume in which the two conditions
are satisfied is of order $\delta s^2$, except when $Q$ is itself
of order $\delta s$ or superior. In this case the two shells
coincide with a precision at least of order $\delta s$, and the
integration volume is of order $\delta s$. In the limit in which
$\delta s \ra 0$, the diagram will be null unless $Q = 0$. Then,
if $\bfk = k \,\hat \Omega$,  we can rewrite Eq. (\ref{volume}) as
\bea \delta_{Q;0} \Lambda^d \Omega_d \; G_1(\Lambda) \;
G_2(\Lambda)  \D\int \D\frac{d\Omega_d}{\Omega_d}  \; f(\bfq_1,
-\bfq_1, \bfq_3, -\bfq_3,  \Lambda \hat \Omega, -\Lambda \hat
\Omega) \;  \delta s. \ena Diagrams with more than two vertices
can be analyzed in similar terms.

There are two other rules for the diagrams of the first type. For
each internal line connecting fields at times $t$ and $t'$, there
will be a factor $\cG(\Lambda, t,t')$ for contractions of the type
$\lt<\cp\cp\rt>$, or $-i\mG(\Lambda, t, t')$ in the case of
$\lt<\cp \sqp\rt>$. For each internal line there will be two time
integrals between $0$ and $T$, associated with the times of the
two fields of type $>$ joined by the line.

Diagrams of the second type, shown in Fig. \ref{type 2}, consist
of a single chain of propagators, all carrying the same momentum
$\bfk$, with $|\bfk| = \Lambda$. For each internal line connecting
fields at times $t$ and $t'$, there will be a factor $\cG(\Lambda,
t,t')$ or $\mG(\Lambda, t, t')$. For each intermediate  vertex
(i.e., not at the extremes of the chain) with external lines
carrying total momentum equal to $\bfq$, there will be a Kronecker
delta $\delta_{q; 0}$ (the argument is essentially the same that
was given above for diagrams of the first type). Finally, if the
total external momentum entering to the diagram at one of its
extremes is $\bfk$, there will be a factor $\Lambda \, \delta(k -
\Lambda)$. As before, there will be two time integrals for each
internal line.

Note that in both types of diagrams, two internal lines at most
are attached to each vertex.

\section{\label{appendix propagators} Propagators}

The propagators obtained for the free action $S_0$ of Eq.(\ref{the
action 2}), with initial conditions given by Eq. (\ref{initial
density matrix}), are defined as in Eqs. (\ref{propagador 1}) and
(\ref{propagador 2}), with \bea \label{prop mG} \mG(k, t, t') =
\D\frac{2}{\omega} \sin\lt[\omega (t - t')\rt]  \, e^{-\kappa
(t-t')} \; \,\Theta(t-t'), \ena and \bea  &\cG(k, t, t') =&  \nn \\ \nn \\
&\D\frac{2}{a} e^{-\kappa (t+t')} \, \lt\{[1+2f(a\,\beta)] -
\D\frac{a \nu}{2 \kappa \omega_0^2} \rt\}
\lt\{\D\frac{\omega_0^2}{\omega^2} \cos[\omega (t-t')] -
\D\frac{\kappa^2}{\omega^2} \cos[\omega(t+t')]
+ \D\frac{\kappa}\omega \sin[\omega (t+t')] \rt\} &\nn \\ \nn \\
&+\D\frac{\nu}{\kappa \omega_0^2} \bigg(\lt\{\cos[\omega(t-t')] +
\D\frac{\kappa}{\omega} \sin[\omega (t-t')]\rt\} e^{-\kappa
(t-t')}\, \Theta(t-t') + (t\leftrightarrow t') \bigg) &  \nn \\ \nn \\
&+\ D\frac{2}{a} \lt(\D\frac{a^2-\omega_0^2}{\omega^2}\rt)  \,
[1+2f(a\, \beta)] \lt\{\cos[\omega(t-t')] -
\cos[\omega(t+t')]\rt\} \,e^{-\kappa (t+t')} ,& \label{prop
cG}\ena  with \bea \omega_0^2 &=& k^2 + m^2, \\  \omega^2 &=&
\omega_0^2 - \kappa^2, \ena and \bea \label{efe} f(x) =
(e^{x}-1)^{-1}.\ena

\section{Obtaining $\delta S$ \label{obtaining delta S}}

Here we will sketch how to compute the actual values of $\eta$,
$\delta \kappa$, etc. from Eq. (\ref{delta S 2}). We will follow
the procedure presented in Appendix \ref{section scalar field
considered}. To illustrate the method, we will give some examples
of diagram calculations, and show how to extract from $\delta S$
the corrections to the parameters in the free action and the value
of $\eta$.

According to the definitions (\ref{division 2}) and (\ref{the
action 3}), and due to the quadratic nature of $S_0$, it results
\bea \Delta S[\sqp_<, \cp_<, \sqp_>, \cp_>] = S_0[\sqp_>, \cp_>] +
S_{\rm int}[\sqp_<+\sqp_>, \cp_<+\cp_>] - S_{\rm int}[\sqp_<,
\cp_<]. \ena Then, from definition (\ref{def SI}), it is \bea
\label{example SI} S_I[\sqp_<, \cp_<, \sqp_>, \cp_>] = S_{\rm
int}[\sqp_<+\sqp_>, \cp_<+\cp_>] - S_{\rm int}[\sqp_<, \cp_<].\ena
We will take as an example the term $v_{41} \;  \sqp \cp^3$.
Writing $\sqp = \sqp_<+\sqp_>$ and $\cp = \cp_<+ \cp_>$, it yields
\bea & \sqp_1 \cp_2 \cp_3 \cp_4 = \sqp_{1>} \cp_{2>} \cp_{3>}
\cp_{4>} + 3 \; \sqp_{1>} \cp_{2>} \cp_{3>} \cp_{4<}+ \sqp_{1<}
\cp_{2>} \cp_{3>} \cp_{4>}+ 3 \;\sqp_{1>} \cp_{2>} \cp_{3<}
\cp_{4<} &\nn  \\ \label{example 1} & +3 \;\sqp_{1<} \cp_{2<}
\cp_{3>} \cp_{4>} + \sqp_{1>} \cp_{2<} \cp_{3<} \cp_{4<}+ 3 \;
\sqp_{1<} \cp_{2<} \cp_{3<} \cp_{4>} + \sqp_{1<} \cp_{2<} \cp_{3<}
\cp_{4<}.&  \ena Here $\sqp_{1<}$ stands for $\sqp_<(\bfk_1,
t_1)$, and so on. The fact that the function $v_{41}$ is symmetric
with respect to the variables corresponding to the $\cp$ fields
accounts for the factors 3. To write $S_I$, the last term in Eq.
(\ref{example 1}), which corresponds to the term $S_{\rm
int}[\sqp_<, \cp_<]$ in Eq. (\ref{example SI}), must be taken
apart. Moreover, since, as it was said before, at most two
internal lines can be attached to any vertex, we can discard the
terms in Eq. (\ref{example 1}) with more than two fields of the
type $>$. Hence, corresponding to $\sqp \cp^3$ in $S_{\rm int}$,
$S_I$ will display the following terms \bea & \Omega_d^{-1}
\lt(\D\prod_{i=1}^4 \D\int_0^T dt_i  \D\int_\Lambda d^d\!k_i\rt)
\, \delta^d\lt(\bfk_1+\dots+\bfk_4\rt) \, \Big(3 \; \sqp_{1>}
\cp_{2>} \cp_{3<} \cp_{4<}+ 3 \; \sqp_{1<} \cp_{2<} \cp_{3>}
\cp_{4>}  &\nn\\ \nn \\  & \label{example terms} + \sqp_{1>}
\cp_{2<} \cp_{3<} \cp_{4<}+ 3\; \sqp_{1<} \cp_{2<} \cp_{3<}
\cp_{4>}\Big) v_{41}(\{\bfk\}; \{t\}).&  \ena Using solely these
terms, we can construct two diagrams with one vertex (shown in
Fig. 6) and which correspond to the term $\lt<S_I\rt>$ in Eq.
(\ref{delta S 2}), and four diagrams with two vertices (shown in
Fig. 7)  corresponding to the term $\lt<S_I^2\rt>$ in Eq.
(\ref{delta S 2}).

Since all the propagators will be evaluated at $\Lambda$, we will
write $\mG(t,t')$ for $\mG(\Lambda, t, t')$, and so for $\cG$.


\vspace{0.8cm}


\begin{center}

\begin{fmffile}{diag2}

\begin{tabular}{|cc|cc|}

\hline & & & \\ & & & \\

\hspace{1cm}
 \begin{fmfgraph*}(3.9,2.4)
   \fmfbottomn{i}{2}
   \fmf{dashes,label=$$,label.side=left}{i1,v1}
   \fmf{plain,label=$$,label.side=left}{v1,i2}
   \fmf{plain,tag=1,label=$\cG$,tension=1/2}{v1,v1}
   \fmffreeze
   \fmfdot{v1}
   \fmflabel{$v_{41}$}{v1}
   \fmflabel{$\sqp_<$}{i1}
   \fmflabel{$\cp_<$}{i2}
 \end{fmfgraph*}

&  &  \hspace{1.1cm}

\begin{fmfgraph*}(3.9,2.4)
   \fmfbottomn{i}{2}
   \fmf{plain,label=$$,label.side=left}{i1,v1}
   \fmf{plain,label=$$,label.side=left}{v1,i2}
   \fmf{phantom,tag=1,label=$-i\mG$,tension=1/2}{v1,v1}
   \fmffreeze
   \fmfi{plain}{subpath (0,1) of vpath1(__v1,__v1)}
   \fmfi{dashes}{subpath (1,2) of vpath1(__v1,__v1)}
   \fmfdot{v1}
   \fmflabel{$v_{41}$}{v1}
   \fmflabel{$\cp_<$}{i1}
   \fmflabel{$\cp_<$}{i2}
\end{fmfgraph*}

& \hspace{0.9cm} \\ & & &

\poi \\\hspace{0.5cm}  Diagram 1.I

& \hspace{1.1cm} & \hspace{0.5cm}

Diagram 1.II

& \\ & & &  \\ \hline

\end{tabular}
\end{fmffile}
\end{center}


\vspace{.5cm}

\noindent {\footnotesize FIG. 6. One-vertex diagrams constructed
using the first two terms in Eq. (\ref{example terms}). Continuous
(dashed) lines represent $\cp$ ($\sqp$) fields.}

\begin{center}

\begin{fmffile}{diag3}

\begin{tabular}{|cc|cc|}

\hline & & & \\ & & & \\

\hspace{1cm}
\begin{fmfgraph*}(3.9,2.4)
   \fmfleftn{i}{2}
   \fmfrightn{o}{2}
   \fmf{plain,label=$ $,label.side=left}{i1,v1}
   \fmf{plain,label=$ $,label.side=right}{o1,v2}
   \fmf{dashes,label=$ $,label.side=right}{i2,v1}
   \fmf{plain,label=$ $,label.side=left}{o2,v2}
   \fmf{phantom,left,tag=1,label=$-i\mG$,tension=1/2}{v1,v2}
   \fmf{phantom,left,tag=1,label=$\cG$,tension=1/2}{v2,v1}
   \fmffreeze
   \fmfi{plain}{subpath (0,1) of vpath1(__v1,__v2)}
   \fmfi{dashes}{subpath (1,2) of vpath1(__v1,__v2)}
   \fmfi{plain}{subpath (0,1) of vpath1(__v2,__v1)}
   \fmfi{plain}{subpath (2,1) of vpath1(__v2,__v1)}
   \fmfdot{v1,v2}
   \fmflabel{$v_{41}$}{v1}
   \fmflabel{$v_{41}$}{v2}
   \fmflabel{$\cp_{<2}$}{i1}
   \fmflabel{$\sqp_{<1}$}{i2}
   \fmflabel{$\cp_{<4}$}{o1}
   \fmflabel{$\cp_{<3}$}{o2}
\end{fmfgraph*}

& \hspace{1cm}&  \hspace{1cm}

\begin{fmfgraph*}(3.9,2.4)
   \fmfleftn{i}{2}
   \fmfrightn{o}{2}
   \fmf{plain,label=$ $,label.side=left}{i1,v1}
   \fmf{plain,label=$ $,label.side=right}{o1,v2}
   \fmf{dashes,label=$ $,label.side=right}{i2,v1}
   \fmf{dashes,label=$ $,label.side=left}{o2,v2}
   \fmf{phantom,left,tag=1,label=$\cG$,tension=1/2}{v1,v2}
   \fmf{phantom,left,tag=1,label=$\cG$,tension=1/2}{v2,v1}
   \fmffreeze
   \fmfi{plain}{subpath (0,1) of vpath1(__v1,__v2)}
   \fmfi{plain}{subpath (1,2) of vpath1(__v1,__v2)}
   \fmfi{plain}{subpath (0,1) of vpath1(__v2,__v1)}
   \fmfi{plain}{subpath (2,1) of vpath1(__v2,__v1)}
   \fmfdot{v1,v2}
   \fmflabel{$v_{41}$}{v1}
   \fmflabel{$v_{41}$}{v2}
   \fmflabel{$\cp_{<3}$}{i1}
   \fmflabel{$\sqp_{<1}$}{i2}
   \fmflabel{$\cp_{<4}$}{o1}
   \fmflabel{$\sqp_{<2}$}{o2}
\end{fmfgraph*}

& \hspace{1cm} \\ & & &

\poi \\\hspace{0.5cm}  Diagram 2.I

& \hspace{0.8cm} & \hspace{0.5cm}

Diagram 2.II

& \hspace{0.5cm}  \\ & & &  \\ \hline

& & & \\ & & & \\


\hspace{1cm}
\begin{fmfgraph*}(3.9,2.4)
   \fmfleftn{i}{2}
   \fmfrightn{o}{2}
   \fmf{plain,label=$ $,label.side=left}{i1,v1}
   \fmf{plain,label=$ $,label.side=right}{o1,v2}
   \fmf{plain,label=$ $,label.side=right}{i2,v1}
   \fmf{plain,label=$ $,label.side=left}{o2,v2}
   \fmf{phantom,left,tag=1,label=$-i\mG$,tension=1/2}{v1,v2}
   \fmf{phantom,left,tag=1,label=$-i\mG$,tension=1/2}{v2,v1}
   \fmffreeze
   \fmfi{plain}{subpath (0,1) of vpath1(__v1,__v2)}
   \fmfi{dashes}{subpath (1,2) of vpath1(__v1,__v2)}
   \fmfi{plain}{subpath (0,1) of vpath1(__v2,__v1)}
   \fmfi{dashes}{subpath (2,1) of vpath1(__v2,__v1)}
   \fmfdot{v1,v2}
   \fmflabel{$v_{41}$}{v1}
   \fmflabel{$v_{41}$}{v2}
   \fmflabel{$\cp_{<2}$}{i1}
   \fmflabel{$\cp_{<1}$}{i2}
   \fmflabel{$\cp_{<4}$}{o1}
   \fmflabel{$\cp_{<3}$}{o2}
\end{fmfgraph*}

& \hspace{1cm}&  \hspace{1cm}

\begin{fmfgraph*}(3.9,2.5)
   \fmfleftn{i}{3}
   \fmfrightn{o}{3}
   \fmf{plain}{i1,v1}
   \fmf{plain,label=$v_{41}$,label.side=left}{i2,v1}
   \fmf{dashes}{i3,v1}
   \fmf{plain}{o1,v2}
   \fmf{plain,label=$v_{41}$,label.side=right}{o2,v2}
   \fmf{plain}{o3,v2}
   \fmf{phantom,straight,tag=1,label={\footnotesize$-i\mG$}}{v1,v2}
   \fmffreeze
   \fmfi{plain}{subpath (0,1/3) of vpath1(__v1,__v2)}
   \fmfi{dashes}{subpath (1/3,1) of vpath1(__v1,__v2)}
   \fmflabel{$\sqp_{<1}$}{i3}
   \fmflabel{$\cp_{<2}$}{i2}
   \fmflabel{$\cp_{<3}$}{i1}
   \fmflabel{$\cp_{<6}$}{o1}
   \fmflabel{$\cp_{<5}$}{o2}
   \fmflabel{$\cp_{<4}$}{o3}
   \fmfdot{v1,v2}
\end{fmfgraph*}

& \hspace{1cm} \\ & & &

\poi \\\hspace{0.5cm}  Diagram 2.III

& \hspace{0.5cm} & \hspace{0.5cm}

Diagram 2.IV

& \\ & & &  \\ \hline

\end{tabular}
\end{fmffile}
\end{center}

\noindent {\footnotesize FIG. 7. Two-vertex diagrams constructed
using the last two terms in Eq. (\ref{example terms}). Continuous
(dashed) lines represent $\cp$ ($\sqp$) fields.}


\poi We use continuous lines for $\cp$ fields and dashed for
$\sqp$. In this way, internal lines corresponding to propagators
$\cG$ are represented by continuous lines, and corresponding to
propagators $-i\mG$ by half dashed, half continuous lines. We use
$\sqp_{<i}$ for $\sqp(\bfk_i, t_i)$ and so on. Now we analyze each
diagram in separated subsections.

\subsection{Diagram 1.I}

The first one-vertex diagram in Fig. 6 comes from the term $3
\sqp_< \cp_< \cp_> \cp_>$ in Eq. (\ref{example terms}), after
contracting the last two $\cp_>$ fields. There just one way in
which this contraction can be made, so there is no additional
combinatorial factors. According to the rules enunciated at the
end of Sec. \ref{section perturbative approach}, there would be a
Kronecker delta $\delta_{|\bfk_1+\bfk_2|;0}$, but the condition
$|\bfk_1+\bfk_2| = 0$ is already granted by the Dirac delta of
momentum conservation. Thus, in this case the Kronecker delta can
be omitted. The contribution of this diagram to $\delta S$ reads
\bea \label{diagram I} \D\int_{0}^{T}  dt_1 \int_0^T dt_2
\D\int_{b\Lambda} d^d\!k \, \sqp(\bfk, t_1) \cp(-\bfk, t_2) \;
v(k; t_1, t_2) \delta s, \ena where \bea v(k; t_1, t_2) = 3 \D\int
\D\frac{d\Omega_d}{\Omega_d} \D\int_0^T dt_3 \int_0^T dt_4\;
v_{41}(\bfk, -\bfk, \hat \Omega,-\hat \Omega; t_1, t_2, t_3, t_4)
\cG(t_3, t_4). \ena We have written $v \, \delta s$ and not
directly something like $\delta v^{{\rm(1.I)}}_{21}$, because this
is not yet the actual contribution of Diagram 1.I to $\delta
v_{21}$ . We have to remove from $v$ those terms associated with
$\eta$, $\delta m^2$ and $\delta \kappa$. We must separate in
$v(k; t_1, t_2)$ something proportional to $\delta(t_1-t_2)$ and
something proportional to $\partial \delta(t_1-t_2)/ t_2$. Two
projectors are introduced. Given a function of two times $v(k;,
t_1, t_2)$, we define \bea \label{bfP} \bfP v(k; t_1, t_2) &=& \cP
v(k) \; \delta(t_1-t_2), \ena and, if $v(k; t_1, t_2) = 0$ for
$t_2 > t_1$, \bea \label{bfQ} \bfQ v(k; t_1, t_2) &=& \cQ v(k) \;
\lt[2 \lt(\D\frac{\partial }{\partial t_2} + \delta(t_2) -
\delta(0)\rt) \delta(t_1-t_2)\rt], \ena where \bea \label{def cP}
\cP v(k) &=& \D\frac1T \int_0^T dt_1 \int_0^T dt_2 \; v(k; t_1,
t_2), \ena and \bea \label{def cQ} \cQ v(k) &=& \D\frac1T \int_0^T
dt_1 \int_0^T dt_2 \; v(k; t_1, t_2) \; (t_2-t_1). \ena It is easy
to verify that $\bfP^2 = \bfP$, $\bfQ^2 = \bfQ$, and that $\bfQ
\bfP = \bfP \bfQ = 0$. This proves that the decomposition \bea
\label{decomposition} v(k; t_1, t_2) = \bfP v(k; t_1, t_2) + \bfQ
v(k; t_1, t_2) + \Delta v(k; t_1, t_2) \ena is unique. Defining
\bea  v_0 = \cP v, \label{definicion v0} \ena and \bea
\label{definicion v1} v_1 = \cQ v, \ena and using
(\ref{decomposition}) it yields \bea &\D\int_0^T dt_1 \int_0^T
dt_2 \D\int_{b\Lambda} d^d\!k  \;
\sqp(\bfk, t_1) \cp(-\bfk, t_2) \, v(k; t_1, t_2) = & \nn \\ \nn \\
\nn & \D\int_0^T dt_1 \D\int_{b\Lambda} d^d\!k \; \sqp(\bfk, t_1)
\cp(-\bfk, t_1) \, v_0(k) + \D\int_0^T dt_1 \int_0^T dt_2
\D\int_{b\Lambda} d^d\!k \; \sqp(\bfk, t_1) \dot \cp(-\bfk, t_1)
\, v_1(k)  & \nn
\\ \nn \\ &+\D\int_0^T dt_1 \int_0^T dt_2 \D\int_{b\Lambda}
d^d\!k \; \sqp(\bfk, t_1) \cp(-\bfk, t_2) \, \Delta v(k; t_1,
t_2). & \ena  [Observe that the extra delta factors in (\ref{bfQ})
get rid of the boundary terms, and that the factor 2 appears
because the deltas are evaluated at the extreme of integration.]
In this way, we have extracted from $v(k; t_1, t_2)$ two
quantities: $v_0(k)$ which acts as a momentum dependent mass
squared term, and $v_1(k)$ which is equivalent to a damping
constant. Indeed, comparing with (\ref{S0 prima}), we get \bea
\delta \kappa^{{\rm (1.I)}} = - v_1(k) \; \delta s. \ena From
$v_0(k)$ we can extract contributions to $\delta m^2$ and $\eta$.
Writing \bea \label{Delta v0} v_0(k) = v_0(0) + k \D\frac{\partial
v_0(0)}{\partial k} + \D\frac{k^2}{2!} \D\frac{\partial^2
v_0(0)}{\partial k^2} + \Delta v_0(k), \ena matching terms with
Eq. (\ref{S0 prima}), we identify \bea \delta m^{{\rm2(1.I)}} = -2
v_0(0) \; \delta s \ena and \bea \eta^{{\rm(1.I)}} = -\D\frac12
\D\frac{\partial^2 v_0(0)}{\partial k^2}. \ena After subtracting
from $v(k; t_1, t_2)$ all the terms belonging to the free action,
the net contribution from Diagram (1.I) to $\delta v_{21}(k; t_1,
t_2)$ comes from Eqs. (\ref{decomposition}) and (\ref{Delta v0}),
and is \bea \delta v_{21}^{{\rm(1.I)}}(k; t_1, t_2) &=& \Delta
v(k; t_1, t_2) + \Delta v_0(k) \; \delta(t_1-t_2). \ena If $v(k;
t_1, t_2)$ were local in time but not constant, that is, if it
were $v(k; t_1) \delta(t_1-t_2)$, it could be interpreted as a
time dependent mass. Our definition in this case would give for
$\delta m^{{\rm2(1.I)}}$ the time average of $v(0; t_1) \, \delta
s$. The definition of $\bfP$ can be generalized for functions
$g(k; t)$ depending on just one time, writing \bea \bfP g(k; t)
&=& \cP g(k), \ena with \bea \cP g(k) = \D\frac1T \int_0^T dt \;
g(k; t). \ena

Something  analogous can be said about corrections of the form
\bea  i \D\int_{0}^{T}  dt_1 \int_0^T dt_2 \D\int_{b\Lambda}
d^d\!k \, \sqp(\bfk, t_1) \sqp(-\bfk, t_2) \; w(k; t_1, t_2)
\delta s. \ena We have to subtract from $w \,\delta s$ the part
that corresponds to the term \bea \delta\nu\; \sqp(\bfk, t)
\sqp(-\bfk, t) \ena in  (\ref{S0 prima}). In this case is \bea
\delta \nu = 2 w_0(\bfk) \delta s\ena where $w_0 = \cP w$.

\subsection{Diagram 1.II}

Diagram 1.II in Fig. 6 comes from the term $3 \sqp_> \cp_> \cp_<
\cp_<$ in Eq. (\ref{example terms}) after contracting the first
two fields. As could be suspected, it is zero. Terms with no
$\sqp$ fields are not included in the definition of the action
[Eq. (\ref{the action 3}) and below], and owing to the causal
nature of the propagator $\lt<\cp \sqp\rt>$,  Eq. (\ref{propagador
2}), they cannot be generated . Remember that we have defined the
action in Sec. \ref{section the action} in such a way that the
there is always at least one field $\sqp$ (let us call it
$\sqp^*$) evaluated at a time $t^*$ greater or equal than the
times of all the $\cp$ fields. When any of the $\cp(\bfk, t)$
fields is paired with $\sqp^*$, there will be a $\Theta(t-t^*)$,
that means that $t$ must be greater than $t^*$. The two conditions
are mutually exclusive and the resulting diagram is effectively
zero. In the case of the vertex $v_{41}$ there is just one $\sqp$,
which must be $\sqp^*$ necessarily.

This is essentially the mechanism which made all the diagrams with
no external $\sqp$ fields equal to zero.

\subsection{Diagram 2.I}

The first diagram in Fig. 7 contributes to $\delta v_{41}$. It has
a combinatorial factor equal to $3$ (ways of choosing which one of
the $\cp$ in the left vertex is $\cp_<$) $\times$ $3$ (ways of
choosing the $\cp_>$ in the right vertex) $\times$ $2$ (ways of
pairing the fields $(\cp_>, \cp_>)$ on the left with the fields
$(\sqp_>, \cp_>$) on the right) $\times$ $2$ (combinatorial factor
from the expansion of $S_I^2$). There is also a factor $-i^2/2!$,
which comes from the $i/2$ in Eq. (\ref{delta S 2}) and the $-i$
in $-i\mG$. The result is \bea &\delta
v_{41}^{{\rm(2.I)}}(\{\bfk\}; \{t\}) = 18  \, \delta s \, \Bigg[
\delta_{|\bfk_1 + \bfk_2|; 0} \D\int_0^T dt_1' \dots \int_0^T
dt_4' \D\int \D\frac{d\Omega_d}{\Omega_d}  v_{41}(\bfk_1, \bfk_2,
\hat \Omega, -\hat \Omega; t_1, t_2, t_3', t_4') &\nn \\ \nn
\\ & \times \mG(t_3', t_1') \, \cG(t_4', t_2')  \; v_{41}(\hat \Omega,
-\hat \Omega, \bfk_3, \bfk_4; t_1', t_2', t_3,
t_4)\Bigg]_{\overline{234}}, &\ena where the subscript
$\overline{ij\dots}$ means symmetrization with respect to the
given variables (remember that according to our definitions,
$v_{nm}$ is symmetrical with respect to the permutations of the
$m$ fields of type $\sqp$, and of the $n-m$ of type $\cp$).

\subsection{Diagram 2.II}

The second diagram in Fig. 7 contributes to $\delta v_{42}$, has a
combinatorial factor equal to $3 \times 3 \times 2$ and it gives
\bea & \delta v_{42}^{{\rm(2.II)}}(\{\bfk\}; \{t\}) = 9 \, \delta
s \, \Bigg[ \delta_{|\bfk_1 + \bfk_3|; 0}   \D\int_0^T dt_1' \dots
\int_0^T dt_4' \D\int \D\frac{d\Omega_d}{\Omega_d}  v_{41}(\bfk_1,
\bfk_3, \hat \Omega, -\hat \Omega; t_1, t_3, t_1', t_2') &\nn \\
\nn
\\ & \times \cG(t_1', t_3') \, \cG(t_2', t_4')  \; v_{41}(\bfk_2,
\bfk_4, \hat \Omega, -\hat \Omega; t_2, t_4, t_3',
t_4')\Bigg]_{\overline{12} \; \overline{34}}. &\ena

\subsection{Diagram 2.III}

The third diagram in Fig. 7 is zero due to causality, which
prevents diagrams with no external $\sqp$ fields.

\subsection{Diagram 2.IV \label{diagram 2.IV}}

The last diagram in Fig. 7 is an example of a tree diagram. It
contributes to $\delta v_{61}$, is proportional to
$\delta(|\bfk_1+\bfk_2+\bfk_3|-1)$, and has a combinatorial factor
equal to 3, which is the number of ways of choosing $\cp_>$ among
the 3 $\cp$ fields [see Eq. (\ref{example terms})]. It results
\bea &\delta v_{61}^{{\rm(2.IV)}}(\{\bfk\}; \{t\})\! =\! 3\!
\bigg[\delta(|\bfk_1+\bfk_2+\bfk_3|\!-\!1) \! \D\int_0^T \!\!\!
dt_1' \! \int_0^T  \!\!\!dt_4' \, v_{41}(\bfk_1, \bfk_2, \bfk_3,
-\bfk_1-\!\bfk_2-\!\bfk_3; t_1, t_2, t_3, t_4') & \nn \\   &
\times \mG(t_4', t_1')\, v_{41}(\bfk_1+\bfk_2+\bfk_3,\, \bfk_4,
\bfk_5, \bfk_6; t_1', t_2, t_3, t_4)\bigg]_{\overline{23456}} .&
\ena

\section{\label{appendix rescaling} Rescaling}

To see in detail the effect of the rescaling, take a generic term
$v_{nm}$ in the interaction part (terms in the free action can be
analyzed in the same way). After step i) has been performed the
corresponding term in the action is (constant factors omitted)
\bea &\D\int_0^T dt_1 \dots \D\int_0^T dt_n \D\int_{b\Lambda}
d^d\!k_1 \dots \D\int_{b\Lambda} d^d\!k_n \;
\delta^d(\bfk_1+\dots+\bfk_n)& \nn \\ & \times v'_{nm}(\bfk_1,
\dots, \bfk_n; t_1, \dots, t_n) \; \lt[\sqp(\bfk_1, t_1) \dots
\sqp(\bfk_m, t_m)\rt] \lt[\cp(\bfk_{m+1}, t_{m+1}) \dots
\cp(\bfk_n, t_n)\rt], &\ena where  $v_{nm}'$ is given in Eq.
(\ref{vnm prima}). Then, redefine the fields and change
integration variables according to Eqs. (\ref{resc 1})-(\ref{resc
4}). The momentum integrals are again restricted to $k\le\Lambda =
1$, but the time interval goes up to $b^{\alpha_t}\,T$: \bea
&\D\int_0^{b^{\alpha_t} T} dt_1 \dots \D\int_0^{b^{\alpha_t} T}
dt_n \D\int_{\Lambda} d^d\!k_1 \dots \D\int_{\Lambda} d^d\!k_n \;
\delta^d(\bfk_1+\dots+\bfk_n) \; b^{d(n-1) - n \alpha_t + n
\alpha_\sqp} \;\, \nn&  \\& \times v'_{nm}(b \bfk_1, \dots;
b^{-\alpha_t} t_1, \dots) \; \lt[\sqp(\bfk_1, t_1) \dots \rt]
\lt[\cp(\bfk_{m+1}, t_{m+1}) \dots \rt]. \label{rescaling 1} &
\ena We expand $v'_{nm}(b \bfk_1, \dots, b \bfk_n; b^{-\alpha_t}
t_1, \dots, b^{-\alpha_t} t_n)$ around $\bfk_i$ and $t_i$, and the
time integrals as functions of their superior extreme around $T$.
To order $\delta s$ we get \bea & \D\int_0^T dt_1 \dots
\D\int_0^{T} dt_n \D\int_{\Lambda} d^d\!k_1 \dots \D\int_{\Lambda}
d^d\!k_n \; \delta^d(\bfk_1+\dots+\bfk_n) & \nn
\\   &\times \lt[1 + \lt\{-d(n-1) + n \alpha_t - n \alpha_\sqp -
\bfk_i \D\frac{\partial}{\partial \bfk_i} + \alpha_t \, t_i\,
\D\frac{\partial}{\partial t_i} \rt\} \delta s\rt] v'_{nm}(\bfk_1,
\dots; t_1, \dots)& \nn \\ \nn & \times  \lt[\sqp(\bfk_1, t_1)
\dots \rt] \lt[\cp(\bfk_{m+1}, t_{m+1}) \dots \rt]& \ena \bea &
-\alpha_t T \; \delta s \, \D\sum_{j=1}^n \, \D\int_0^T  dt_1
\dots \D\int_0^{T} dt_n \D\int_{\Lambda} d^d\!k_1 \dots
\D\int_{\Lambda} d^d\!k_n \; \delta^d(\bfk_1+\dots+\bfk_n) & \nn
\\ &\times 2\delta(t_j-T) \; v'_{nm}(\bfk_1, \dots; t_1, \dots)
\;\; \lt[\sqp(\bfk_1, t_1) \dots \rt] \lt[\cp(\bfk_{m+1}, t_{m+1})
\dots \rt].& \ena Because of the $\delta(t_j-T)$, the last two
lines are a sum of terms with $n-1$ time integrals. One field (the
one with index $j$) in each term of the sum is evaluated at time
$T$. Thus, in principle boundary terms could appear in the action.
But this is not the case: if $j \le m$, the field evaluated at
$t_j = T$ is of type $\sqp$, and the corresponding term is zero,
because $\sqp(\bfk, T) = 0$ by the CTP condition (\ref{CTP
condition 2}). If $j\ge m $ then the field evaluated at the
boundary if of type $\cp$. But we have imposed the condition that
in each term of the action there is always a field $\sqp$
evaluated at a time $t^*$ which is equal or greater than the times
at which are evaluated all the $\cp$ fields. Because $t_j = T$,
$t^*$ must be equal to $T$. But, again, the CTP condition implies
that the term will be zero. Thus, we could rewrite Eq.
(\ref{rescaling 1}) keeping $T$ as the superior limit of the time
integrals, which means that $T$ can be considered as an invariant
quantity with respect to the RG transformation.

In conclusion, rescaling takes  the resulting parameters after
mode elimination, Eqs. (\ref{S0 prima}) and (\ref{vnm prima}), and
returns (to order $\delta s$) \bea \label{ec diferencias m2} & m^2
+ \delta m^2 + \Big(-2\alpha_\sqp +
\alpha_t - d  \Big) m^2 \; \delta s,& \\
\nn \\  &\kappa + \delta \kappa + \lt(-2\alpha_\sqp
- d - k \D\frac{\partial}{\partial k} \rt) \kappa \; \delta s,& \\
\nn \\ & \nu +\delta \nu + \lt(-2\alpha_\sqp + \alpha_t - d
-k\D\frac{\partial}{\partial k}  \rt) \nu \; \delta s,& \\ \nn \\
&v_{nm}+\delta v_{nm} + \lt[- n \alpha_\sqp + n \alpha_t -d(n-1) -
\bfk_i \D\frac{\partial}{\partial \bfk_i} + \alpha_t \, t_i\,
\D\frac{\partial}{\partial t_i}  \rt] v_{nm}\;\delta s. &\ena

Rescaling also affects the density matrix. After rescaling it
results \bea &\rho'[\sqp(\bfk,0), \cp(\bfk, 0)] =
\exp\Bigg\{-\D\int_\Lambda d^d\!k \;b^{d+2\alpha_\sqp} \frac{a(b
k)}{4} \Bigg[ \tanh\lt(\frac{a(b k) \beta(b k)}{2}\rt)\;
\sqp(\bfk, 0) \sqp(-\bfk, 0)  & \nn \\ \nn \\ \nn
&+\coth\lt(\D\frac{a(b k) \beta(b k)}{2}\rt) \;\cp(\bfk, 0)
\cp(-\bfk, 0) \Bigg] \Bigg\} = & \\ \nn \\ \label{rescaled density
matrix 1} &\!\!\!\!\!\!\!\exp\Bigg\{\!\!\!-\!\!\D\int_\Lambda
\!\!d^d\!k \frac{\lt[a(b k) b^{d+2\alpha_\sqp}\rt]}{4} \! \Bigg[\!
\tanh\lt(\frac{\lt[a(b k)b^{d+2\alpha_\sqp}\rt] \!\! \lt[\beta(b
k) b^{-d-2\alpha_\sqp}\rt]}{2}\rt) \! \sqp(\bfk, 0) \sqp(-\bfk, 0)
+ \!\dots \!\Bigg]\!\Bigg\}. & \ena Then, rescaling takes $a$ and
$\beta$, and returns \bea a + \lt(-d - 2\alpha_\phi
- k \D\frac{\partial}{\partial k}\rt) a \; \delta s, \\ \nn \\
\beta + \lt(d + 2\alpha_\phi - k \D\frac{\partial}{\partial k}\rt)
\beta \; \delta s. \label{ec diferencias beta} \ena

\subsection{Final remark}

Before leaving this section, we note a special case of rescaling.
In principle,  one does not have to rescale the arguments of
$\delta \kappa$, $\delta \nu$ and $\delta v_{nm}$, because they
are already of order $\delta s$. Note that there are no
$k$-derivatives nor $t$-derivatives of these quantities in the
differential equations of Sec. \ref{section RG equations and
formal solutions}. But it may happen (for example in the case of
tree diagrams) that \bea \delta v_{nm}(\{\bfk\};\{t\}) \propto
\delta(|\bfK|-\Lambda) \; \delta s, \ena where $\bfK$ is a linear
combination of the $\bfk_i$. Rescaling $\bfk_i$ gives \bea \delta
v_{nm}(b \{\bfk\}; \{t\}) \propto \delta(b |\bfK| -\Lambda) \;
\delta s \ena or

\vspace{-1cm}

\bea \delta v_{nm}(b\{\bfk\}, \{t\}) \propto b^{-1} \,
\delta(|\bfK|-b^{-1}) \; \delta s. \label{delta +} \ena The first
$b^{-1}$ can be replaced by 1, but $b^{-1} \approx 1 + \delta s$
inside the delta must be conserved because it can define how to
take limit values of the functions. We define $\delta (k-1^+)$
such that \bea \label{def delta +}
\int_0^q dk \; \delta(k-\Lambda^+) = \left\{%
\begin{array}{ll}
    0, & \hbox{if $q\le \Lambda$,} \\
    & \\
    1 & \hbox{if $q > \Lambda$} \\
\end{array}%
\right. \ena and then rewrite Eq. (\ref{delta +}) as \bea \delta
v_{nm}(b\{\bfk\}, \{t\}) \propto \delta(|\bfK|- \Lambda^+) \;
\delta s. \ena
\section{\label{appendix V} Closed set of couplings to order $\lambda^2$}

Here we find the set of couplings for which the RG transformation
is closed to order $\lambda^2$. Starting from the usual $\lambda
\sqp^4$ theory, Eq. (\ref{the action 1}),  new terms generated by
the RG, and $v_{41}$ and $v_{43}$ themselves, will not longer be
local in time. After one infinitesimal step, there will be terms
with fields evaluated at two times; for example the one generated
by the diagram in Fig. 8.

\vspace{0.8cm}

\begin{center}
\begin{fmffile}{diag4}
\begin{fmfgraph*}(3.9,2.4)
   \fmfleftn{i}{2}
   \fmfrightn{o}{2}
   \fmf{plain,label=$ $,label.side=left}{i1,v1}
   \fmf{plain,label=$ $,label.side=right}{o1,v2}
   \fmf{dashes,label=$ $,label.side=right}{i2,v1}
   \fmf{plain,label=$ $,label.side=left}{o2,v2}
   \fmf{phantom,left,tag=1,label=$-i\mG$,tension=1/2}{v1,v2}
   \fmf{phantom,left,tag=1,label=$\cG$,tension=1/2}{v2,v1}
   \fmffreeze
   \fmfi{plain}{subpath (0,1) of vpath1(__v1,__v2)}
   \fmfi{dashes}{subpath (1,2) of vpath1(__v1,__v2)}
   \fmfi{plain}{subpath (0,1) of vpath1(__v2,__v1)}
   \fmfi{plain}{subpath (2,1) of vpath1(__v2,__v1)}
   \fmfdot{v1,v2}
   \fmflabel{$\lambda,\, t$}{v1}
   \fmflabel{$\lambda,\, t'$}{v2}
   \fmflabel{$\sqp$}{i2}
   \fmflabel{$\cp$}{i1}
   \fmflabel{$\cp'$}{o1}
   \fmflabel{$\cp'$}{o2}
   \fmfbottom{b}
\end{fmfgraph*}
\end{fmffile}

\end{center}

\noindent {\footnotesize FIG. 8. One-loop diagram drawn using two
identical vertices local in time, with coupling constant
$\lambda$. The resulting term in the effective action will be no
longer local in time, but will couple fields at $t$ ($\sqp \cp$)
and $t'$ ($\cp' \cp'$).}

\vspace{0.5cm}

\noi We used $\sqp$ and $\cp$ for fields evaluated at time $t$,
and $\sqp'$ and $\cp'$ at $t'$. If we iterate the transformation
once more, we will get terms with fields evaluated at 3 and 4
times. We may suppose that, basically, $v_{41}$ and $v_{43}$
remain local and time independent, and that the non local and time
dependent terms are corrections.

We assume that $v_{41}$ and $v_{43}$ are of order $\lambda$ and
compute the RG equations to order $\lambda^2$. We have to find
what kind of couplings can be generated when proceeding up to this
order. We assume that for $s \ge 0$ in the interaction part of the
action there will be two terms local in time, \bea \label{term
V41} \D\int_{0}^{T} \!\! dt \D\int_{b\Lambda} {d^d\!k_1 \dots
d^d\!k_4} \, \delta^d\lt(\bfk_1+\dots+\bfk_4\rt) \; V_{41}(s) \,
\sqp(\bfk_1, t) \cp(\bfk_2, t) \cp(\bfk_3, t) \cp(\bfk_4, t) \ena
and \bea \label{term V43} \D\int_{0}^{T} \!\! dt \D\int_{b\Lambda}
{d^d\!k_1 \dots d^d\!k_4} \, \delta^d\lt(\bfk_1+\dots+\bfk_4\rt)
\; V_{43}(s) \, \sqp(\bfk_1, t) \sqp(\bfk_2, t) \sqp(\bfk_3, t)
\cp(\bfk_4, t), \ena where $V_{41}$ and $V_{43}$ are order
$\lambda$ and do not depend on $\{\bfk\}$, and satisfy \bea
V_{41}(0) = V_{43}(0) = -\D\frac{\Omega_d \lambda}{48 (2\pi)^d}.
\ena Starting from these vertices we perform successive
infinitesimal RG transformations until no new terms are generated.
The finite RG transformation will be closed with respect to the
resulting set of couplings, a desired property which was stressed
before. We do not need to keep trace of the precise value of each
coupling from step to step. At this point we just want to
enumerate the couplings that have to be considered to be
consistent to order $\lambda^2$, and find their general functional
form. Rescaling have to be taken into account only in the deltas
appearing in tree diagrams.

Then, for the first step, we start with $V_{41}$ and $V_{43}$ and
construct all the possible diagrams of order $\lambda$ and
$\lambda^2$, shown in Fig. 9. They all will give new terms. Below
each diagram we have written schematically the corresponding term,
indicating the name of the coupling and the variables on which
depends. We have defined \bea Q_{12\dots} = |\bfk_1 + \bfk_2 +
\dots| .\ena The couplings have, by definition, the same symmetry
that the set of fields that they couple. Most diagrams give a
result that has already the required symmetry.

Next, for the second step, we combine the original terms with the
new ones and construct all possible diagrams of order $\lambda$
and $\lambda^2$, shown in Fig. 10. Only two new terms are
generated at this step, with couplings $W_{21}$ and $W_{22}$, all
of order $\lambda^2$. They cannot be used to produce new terms,
because combined with any other vertex the resulting diagram would
be at least of order $\lambda^3$. Hence, a third step fails to
create new terms. In Fig. 10, we have drawn the diagrams for the
two field couplings at the end, because, in some sense, they are
at the deepest level and depend on all the terms which precede
them. Note that diagram (8) in Fig. 10 will not depend on $k$,
owing to the fact that $W_{41}$ is evaluated at $Q = 0$, and hence
the dependence on $k$ cancels out.


\begin{center}

\begin{fmffile}{diag5}

\begin{tabular}{|cc|cc|cc|}

\hline & & & & & \\ & & & & & \\
\hspace{0.5cm}
  \begin{fmfgraph*}(2.8,1.5)
   \fmfbottomn{i}{2}
   \fmf{dashes,label=$$,label.side=left}{i1,v1}
   \fmf{plain,label=$$,label.side=left}{v1,i2}
   \fmf{plain,tag=1,tension=1/2}{v1,v1}
   \fmffreeze
   \fmfdot{v1}
   \fmflabel{$V_{41}$}{v1}
   \fmflabel{$\sqp$}{i1}
   \fmflabel{$\cp$}{i2}
\end{fmfgraph*}

& \hspace{0.5cm}&  \hspace{0.5cm}

\begin{fmfgraph*}(2.8,1.5)
   \fmfleftn{i}{2}
   \fmfrightn{o}{2}
   \fmf{plain,label=$ $,label.side=left}{i1,v1}
   \fmf{plain,label=$ $,label.side=right}{o1,v2}
   \fmf{dashes,label=$ $,label.side=right}{i2,v1}
   \fmf{plain,label=$ $,label.side=left}{o2,v2}
   \fmf{phantom,left,tag=1,tension=1/2}{v1,v2}
   \fmf{phantom,left,tag=1,tension=1/2}{v2,v1}
   \fmffreeze
   \fmfi{plain}{subpath (0,1) of vpath1(__v1,__v2)}
   \fmfi{dashes}{subpath (1,2) of vpath1(__v1,__v2)}
   \fmfi{plain}{subpath (0,1) of vpath1(__v2,__v1)}
   \fmfi{plain}{subpath (2,1) of vpath1(__v2,__v1)}
   \fmfdot{v1,v2}
   \fmflabel{$V_{41}$}{v1}
   \fmflabel{$V_{41}$}{v2}
   \fmflabel{$\cp$}{i1}
   \fmflabel{$\sqp$}{i2}
   \fmflabel{$\cp'$}{o1}
   \fmflabel{$\cp'$}{o2}
\end{fmfgraph*}

& &

\hspace{0.5cm}
\begin{fmfgraph*}(2.8,1.5)
   \fmfleftn{i}{2}
   \fmfrightn{o}{2}
   \fmf{plain,label=$ $,label.side=left}{i1,v1}
   \fmf{plain,label=$ $,label.side=right}{o1,v2}
   \fmf{dashes,label=$ $,label.side=right}{i2,v1}
   \fmf{dashes,label=$ $,label.side=left}{o2,v2}
   \fmf{phantom,left,tag=1,tension=1/2}{v1,v2}
   \fmf{phantom,left,tag=1,tension=1/2}{v2,v1}
   \fmffreeze
   \fmfi{plain}{subpath (0,1) of vpath1(__v1,__v2)}
   \fmfi{plain}{subpath (1,2) of vpath1(__v1,__v2)}
   \fmfi{plain}{subpath (0,1) of vpath1(__v2,__v1)}
   \fmfi{plain}{subpath (2,1) of vpath1(__v2,__v1)}
   \fmfdot{v1,v2}
   \fmflabel{$V_{41}$}{v1}
   \fmflabel{$V_{41}$}{v2}
   \fmflabel{$\cp$}{i1}
   \fmflabel{$\sqp$}{i2}
   \fmflabel{$\cp'$}{o1}
   \fmflabel{$\sqp'$}{o2}
\end{fmfgraph*}

& \\ & & &  & &

\poi \\ $V_{21}(t) \,\; \sqp  \cp$

& \hspace{0.5cm} & \hspace{0.5cm}

$W_{41}(Q_{12}; t, t') \,\; \sqp \cp \; \cp' \cp'$

& &

$W_{42}(Q_{12}; t, t') \,\; \sqp \cp \; \sqp' \cp'$

& \\ & {\footnotesize (1)}& & {\footnotesize (2)} & &
{\footnotesize (3)} \\ \hline & & & & & \\

\hspace{0.5cm}
  \begin{fmfgraph*}(2.8,1.5)
   \fmfleftn{i}{2}
   \fmfrightn{o}{2}
   \fmf{plain,label=$ $,label.side=left}{i1,v1}
   \fmf{plain,label=$ $,label.side=right}{o1,v2}
   \fmf{dashes,label=$ $,label.side=right}{i2,v1}
   \fmf{dashes,label=$ $,label.side=left}{o2,v2}
   \fmf{phantom,left,tag=1,tension=1/2}{v1,v2}
   \fmf{phantom,left,tag=1,tension=1/2}{v2,v1}
   \fmffreeze
   \fmfi{plain}{subpath (0,1) of vpath1(__v1,__v2)}
   \fmfi{dashes}{subpath (1,2) of vpath1(__v1,__v2)}
   \fmfi{dashes}{subpath (0,1) of vpath1(__v2,__v1)}
   \fmfi{plain}{subpath (2,1) of vpath1(__v2,__v1)}
   \fmfdot{v1,v2}
   \fmflabel{$V_{41}$}{v1}
   \fmflabel{$V_{43}$}{v2}
   \fmflabel{$\cp$}{i1}
   \fmflabel{$\sqp$}{i2}
   \fmflabel{$\cp'$}{o1}
   \fmflabel{$\sqp'$}{o2}
\end{fmfgraph*}

& \hspace{0.5cm} & \hspace{0.5cm}

\begin{fmfgraph*}(2.8,1.5)
   \fmfleftn{i}{2}
   \fmfrightn{o}{2}
   \fmf{plain,label=$ $,label.side=left}{i1,v1}
   \fmf{dashes,label=$ $,label.side=right}{o1,v2}
   \fmf{dashes,label=$ $,label.side=right}{i2,v1}
   \fmf{dashes,label=$ $,label.side=left}{o2,v2}
   \fmf{phantom,left,tag=1,tension=1/2}{v1,v2}
   \fmf{phantom,left,tag=1,tension=1/2}{v2,v1}
   \fmffreeze
   \fmfi{plain}{subpath (0,1) of vpath1(__v1,__v2)}
   \fmfi{dashes}{subpath (1,2) of vpath1(__v1,__v2)}
   \fmfi{plain}{subpath (0,1) of vpath1(__v2,__v1)}
   \fmfi{plain}{subpath (2,1) of vpath1(__v2,__v1)}
   \fmfdot{v1,v2}
   \fmflabel{$V_{41}$}{v1}
   \fmflabel{$V_{43}$}{v2}
   \fmflabel{$\cp$}{i1}
   \fmflabel{$\sqp$}{i2}
   \fmflabel{$\sqp'$}{o1}
   \fmflabel{$\sqp'$}{o2}
\end{fmfgraph*}

& &

\hspace{0.5cm}
\begin{fmfgraph*}(2.8,2.1)
   \fmfleftn{i}{3}
   \fmfrightn{o}{3}
   \fmf{plain}{i1,v1}
   \fmf{plain,label=$V_{41}$,label.side=left}{i2,v1}
   \fmf{dashes}{i3,v1}
   \fmf{plain}{o1,v2}
   \fmf{plain,label=$\;\;V_{41}$,label.side=right}{o2,v2}
   \fmf{plain}{o3,v2}
   \fmf{phantom,straight,tag=1}{v1,v2}
   \fmffreeze
   \fmfi{plain}{subpath (0,1/3) of vpath1(__v1,__v2)}
   \fmfi{dashes}{subpath (1/3,1) of vpath1(__v1,__v2)}
   \fmflabel{$\sqp$}{i3}
   \fmflabel{$\cp$}{i2}
   \fmflabel{$\cp$}{i1}
   \fmflabel{$\cp'$}{o1}
   \fmflabel{$\cp'$}{o2}
   \fmflabel{$\cp'$}{o3}
    \fmfdot{v1,v2}

\end{fmfgraph*}

& \\ & & &  & &

\poi \\ $W_{42}(Q_{12}; t, t') \,\; \sqp \cp \; \sqp' \cp'$

& \hspace{0.5cm} & \hspace{0.5cm}

$W_{43}(Q_{12}; t, t') \,\; \sqp \cp \; \sqp' \sqp'$

& &

{\footnotesize$\;\;\;v_{61}(Q_{123}; t, t') \,\; \sqp \cp \cp \;
\cp' \cp' \cp'$}

& \\  & {\footnotesize (4)} &    & {\footnotesize (5)} &   &
{\footnotesize (6)} \\ \hline & & &  & & \\

\hspace{0.5cm}
\begin{fmfgraph*}(2.8,2.1)
   \fmfleftn{i}{3}
   \fmfrightn{o}{3}
   \fmf{plain}{i1,v1}
   \fmf{plain,label=$V_{41}$,label.side=left}{i2,v1}
   \fmf{dashes}{i3,v1}
   \fmf{plain}{o1,v2}
   \fmf{plain,label=$\;\;V_{41}$,label.side=right}{o2,v2}
   \fmf{dashes}{o3,v2}
   \fmf{phantom,straight,tag=1}{v1,v2}
   \fmffreeze
   \fmfi{plain}{subpath (0,1/3) of vpath1(__v1,__v2)}
   \fmfi{plain}{subpath (1/3,1) of vpath1(__v1,__v2)}
   \fmflabel{$\sqp$}{i3}
   \fmflabel{$\cp$}{i2}
   \fmflabel{$\cp$}{i1}
   \fmflabel{$\cp'$}{o1}
   \fmflabel{$\cp'$}{o2}
   \fmflabel{$\sqp'$}{o3}
   \fmfdot{v1,v2}
\end{fmfgraph*}

& \hspace{0.5cm} & \hspace{0.5cm}

\begin{fmfgraph*}(2.8,2.1)
   \fmfleftn{i}{3}
   \fmfrightn{o}{3}
   \fmf{plain}{i1,v1}
   \fmf{plain,label=$V_{41}$,label.side=left}{i2,v1}
   \fmf{dashes}{i3,v1}
   \fmf{plain}{o1,v2}
   \fmf{dashes,label=$\;\;V_{43}$,label.side=right}{o2,v2}
   \fmf{dashes}{o3,v2}
   \fmf{phantom,straight,tag=1}{v1,v2}
   \fmffreeze
   \fmfi{plain}{subpath (0,1/3) of vpath1(__v1,__v2)}
   \fmfi{dashes}{subpath (1/3,1) of vpath1(__v1,__v2)}
   \fmflabel{$\sqp$}{i3}
   \fmflabel{$\cp$}{i2}
   \fmflabel{$\cp$}{i1}
   \fmflabel{$\cp'$}{o1}
   \fmflabel{$\sqp'$}{o2}
   \fmflabel{$\sqp'$}{o3}
    \fmfdot{v1,v2}
\end{fmfgraph*}

& &

\hspace{0.5cm}
\begin{fmfgraph*}(2.8,2.1)
   \fmfleftn{i}{3}
   \fmfrightn{o}{3}
   \fmf{dashes}{i1,v1}
   \fmf{dashes,label=$V_{43}$,label.side=left}{i2,v1}
   \fmf{dashes}{i3,v1}
   \fmf{plain}{o1,v2}
   \fmf{plain,label=$\;\;V_{41}$,label.side=right}{o2,v2}
   \fmf{plain}{o3,v2}
   \fmf{phantom,straight,tag=1}{v1,v2}
   \fmffreeze
   \fmfi{plain}{subpath (0,1/3) of vpath1(__v1,__v2)}
   \fmfi{dashes}{subpath (1/3,1) of vpath1(__v1,__v2)}
   \fmflabel{$\sqp$}{i3}
   \fmflabel{$\sqp$}{i2}
   \fmflabel{$\sqp$}{i1}
   \fmflabel{$\cp'$}{o1}
   \fmflabel{$\cp'$}{o2}
   \fmflabel{$\cp'$}{o3}
   \fmfdot{v1,v2}
\end{fmfgraph*}

& \\ & & &  & &

\poi \\ {\footnotesize$\;\;\;v_{62}(Q_{123}; t, t') \,\; \sqp \cp
\cp \; \sqp' \cp' \cp'$}

& \hspace{0.5cm} & \hspace{0.5cm}

{\footnotesize$v^{(1)}_{63}(Q_{123}; t, t') \,\; \sqp \cp \cp \;
\sqp' \sqp' \cp'$}

& &

{\footnotesize$\;\;\;v^{(2)}_{63}(Q_{123}; t, t') \,\; \sqp \sqp
\sqp \; \cp' \cp' \cp'$}

& \\  & {\footnotesize (7)} &    & {\footnotesize (8)} & &
{\footnotesize (9)} \\ \hline & & &  & & \\

\hspace{0.5cm}
  \begin{fmfgraph*}(2.8,2.1)
   \fmfleftn{i}{3}
   \fmfrightn{o}{3}
   \fmf{plain}{i1,v1}
   \fmf{plain,label=$V_{41}$,label.side=left}{i2,v1}
   \fmf{dashes}{i3,v1}
   \fmf{dashes}{o1,v2}
   \fmf{dashes,label=$\;\;V_{43}$,label.side=right}{o2,v2}
   \fmf{dashes}{o3,v2}
   \fmf{phantom,straight,tag=1}{v1,v2}
   \fmffreeze
   \fmfi{plain}{subpath (0,1/3) of vpath1(__v1,__v2)}
   \fmfi{plain}{subpath (1/3,1) of vpath1(__v1,__v2)}
   \fmflabel{$\sqp$}{i3}
   \fmflabel{$\cp$}{i2}
   \fmflabel{$\cp$}{i1}
   \fmflabel{$\sqp'$}{o1}
   \fmflabel{$\sqp'$}{o2}
   \fmflabel{$\sqp'$}{o3}
    \fmfdot{v1,v2}
\end{fmfgraph*}
& \hspace{0.5cm} & \hspace{0.5cm}

\begin{fmfgraph*}(2.8,2.1)
   \fmfleftn{i}{3}
   \fmfrightn{o}{3}
   \fmf{dashes}{i1,v1}
   \fmf{dashes,label=$V_{43}$,label.side=left}{i2,v1}
   \fmf{dashes}{i3,v1}
   \fmf{plain}{o1,v2}
   \fmf{dashes,label=$\;\; V_{43}$,label.side=right}{o2,v2}
   \fmf{dashes}{o3,v2}
   \fmf{phantom,straight,tag=1}{v1,v2}
   \fmffreeze
   \fmfi{plain}{subpath (0,1/3) of vpath1(__v1,__v2)}
   \fmfi{dashes}{subpath (1/3,1) of vpath1(__v1,__v2)}
   \fmflabel{$\sqp$}{i3}
   \fmflabel{$\sqp$}{i2}
   \fmflabel{$\sqp$}{i1}
   \fmflabel{$\cp'$}{o1}
   \fmflabel{$\sqp'$}{o2}
   \fmflabel{$\sqp'$}{o3}
   \fmfdot{v1,v2}
\end{fmfgraph*}

& &

\hspace{0.5cm}
\begin{fmfgraph*}(2.8,2.1)
   \fmfleftn{i}{3}
   \fmfrightn{o}{3}
   \fmf{dashes}{i1,v1}
   \fmf{dashes,label=$V_{43}$,label.side=left}{i2,v1}
   \fmf{dashes}{i3,v1}
   \fmf{dashes}{o1,v2}
   \fmf{dashes,label=$\;\;V_{43}$,label.side=right}{o2,v2}
   \fmf{dashes}{o3,v2}
   \fmf{phantom,straight,tag=1}{v1,v2}
   \fmffreeze
   \fmfi{plain}{subpath (0,1/3) of vpath1(__v1,__v2)}
   \fmfi{plain}{subpath (1/3,1) of vpath1(__v1,__v2)}
   \fmflabel{$\sqp$}{i3}
   \fmflabel{$\sqp$}{i2}
   \fmflabel{$\sqp$}{i1}
   \fmflabel{$\sqp'$}{o1}
   \fmflabel{$\sqp'$}{o2}
   \fmflabel{$\sqp'$}{o3}
   \fmfdot{v1,v2}
\end{fmfgraph*}

& \\ & & &  & &

\poi \\ {\footnotesize$\;\;\; v_{64}(Q_{123}; t, t') \,\; \sqp \cp
\cp \; \sqp' \sqp' \sqp'$}

& \hspace{0.5cm} & \hspace{0.5cm}

{\footnotesize$v_{65}(Q_{123}; t, t') \,\; \sqp \sqp \sqp \; \sqp'
\sqp' \cp'$}

& &

{\footnotesize$\;\;\; v_{66}(Q_{123}; t, t') \,\; \sqp \sqp \sqp
\; \sqp' \sqp' \sqp'$}

& \\  & {\footnotesize (10)} &  & {\footnotesize (11)}&   & {\footnotesize (12)} \\
\hline

\end{tabular}
\end{fmffile}
\end{center}

\noindent {\footnotesize FIG. 9. Set of diagrams obtained after
performing one infinitesimal step of RG transformation starting
from the initial condition given by Eq. (\ref{the action 1}).
Below each diagram we indicate the coupling generated, with its
corresponding dependence on the momentum and time variables, with
$Q_{ij\dots} = |\bfk_1+\bfk_2+\dots|$.}


\begin{center}
\begin{fmffile}{diag6}
\begin{tabular}{|cc|cc|cc|}
\hline

& & & & & \\ & & &  & &  \\ & & &  & & \\

\hspace{0.5cm}
  \begin{fmfgraph*}(2.8,1.1)
   \fmfbottomn{i}{4}
   \fmftopn{j}{1}
   \fmf{dashes}{i1,j1}
   \fmf{plain}{j1,i2}
   \fmf{plain}{j1,i3}
   \fmf{plain}{j1,i4}
   \fmf{plain,tag=1,tension=0.7}{j1,j1}
   \fmffreeze
   \fmflabel{$\sqp$}{i1}
   \fmflabel{$\cp$}{i2}
   \fmflabel{$\cp'$}{i3}
   \fmflabel{$\cp'$}{i4}
   \fmflabel{$v_{61}$}{j1}
   \fmfdot{j1}
\end{fmfgraph*}

& \hspace{0.5cm} &  \hspace{0.5cm}

\begin{fmfgraph*}(2.8,1.1)
   \fmfbottomn{i}{4}
   \fmftopn{j}{1}
   \fmf{dashes}{i1,j1}
   \fmf{plain}{j1,i2}
   \fmf{plain}{j1,i3}
   \fmf{plain}{j1,i4}
   \fmf{phantom,tag=1,tension=0.7}{j1,j1}
   \fmffreeze
   \fmfi{dashes}{subpath (0,1) of vpath1(__j1,__j1)}
   \fmfi{plain}{subpath (1,2) of vpath1(__j1,__j1)}
   \fmflabel{$\sqp$}{i1}
   \fmflabel{$\cp$}{i2}
   \fmflabel{$\cp'$}{i3}
   \fmflabel{$\cp'$}{i4}
   \fmflabel{$v_{62}$}{j1}
   \fmfdot{j1}
\end{fmfgraph*}

& &

\hspace{0.5cm}
\begin{fmfgraph*}(2.8,1.1)
   \fmfbottomn{i}{4}
   \fmftopn{j}{1}
   \fmf{dashes}{i1,j1}
   \fmf{plain}{j1,i2}
   \fmf{dashes}{j1,i3}
   \fmf{plain}{j1,i4}
   \fmf{plain,tag=1,tension=0.7}{j1,j1}
   \fmffreeze
   \fmflabel{$\sqp$}{i1}
   \fmflabel{$\cp$}{i2}
   \fmflabel{$\sqp'$}{i3}
   \fmflabel{$\cp'$}{i4}
   \fmflabel{$v_{62}$}{j1}
   \fmfdot{j1}
\end{fmfgraph*}

& \\ & & &  & &

\poi \\ $W_{41}(Q_{12}; t, t') \,\; \sqp \cp \; \cp' \cp'$

& \hspace{0.5cm} & \hspace{0.5cm}

$W_{41}(Q_{12}; t, t') \,\; \sqp \cp \; \cp' \cp'$

& &

$W_{42}(Q_{12}; t, t') \,\; \sqp \sqp \; \cp' \cp'$ & \\

& {\footnotesize (1)}& & {\footnotesize (2)}& & {\footnotesize (3)}\\ \hline & & &  & & \\ & & &  & & \\
& & &  & & \\

\hspace{0.5cm}
  \begin{fmfgraph*}(2.8,1.1)
   \fmfbottomn{i}{4}
   \fmftopn{j}{1}
   \fmf{dashes}{i1,j1}
   \fmf{plain}{j1,i2}
   \fmf{dashes}{j1,i3}
   \fmf{plain}{j1,i4}
   \fmf{phantom,tag=1,tension=0.7}{j1,j1}
   \fmffreeze
   \fmfi{dashes}{subpath (0,1) of vpath1(__j1,__j1)}
   \fmfi{plain}{subpath (1,2) of vpath1(__j1,__j1)}
   \fmflabel{$\sqp$}{i1}
   \fmflabel{$\cp$}{i2}
   \fmflabel{$\sqp'$}{i3}
   \fmflabel{$\cp'$}{i4}
   \fmflabel{$v^{(1)}_{63}$}{j1}
   \fmfdot{j1}
\end{fmfgraph*}

& \hspace{0.5cm} & \hspace{0.5cm}

\begin{fmfgraph*}(2.8,1.1)
   \fmfbottomn{i}{4}
   \fmftopn{j}{1}
   \fmf{dashes}{i1,j1}
   \fmf{plain}{j1,i2}
   \fmf{dashes}{j1,i3}
   \fmf{dashes}{j1,i4}
   \fmf{phantom,tag=1,tension=0.7}{j1,j1}
   \fmffreeze
   \fmfi{plain}{subpath (0,1) of vpath1(__j1,__j1)}
   \fmfi{plain}{subpath (1,2) of vpath1(__j1,__j1)}
   \fmflabel{$\sqp$}{i1}
   \fmflabel{$\cp$}{i2}
   \fmflabel{$\sqp'$}{i3}
   \fmflabel{$\sqp'$}{i4}
   \fmflabel{$v^{(1)}_{63}$}{j1}
   \fmfdot{j1}
\end{fmfgraph*}

& &

\hspace{0.5cm}
\begin{fmfgraph*}(2.8,1.1)
   \fmfbottomn{i}{4}
   \fmftopn{j}{1}
   \fmf{dashes}{i1,j1}
   \fmf{plain}{j1,i2}
   \fmf{dashes}{j1,i3}
   \fmf{dashes}{j1,i4}
   \fmf{phantom,tag=1,tension=0.7}{j1,j1}
   \fmffreeze
   \fmfi{dashes}{subpath (0,1) of vpath1(__j1,__j1)}
   \fmfi{plain}{subpath (1,2) of vpath1(__j1,__j1)}
   \fmflabel{$\sqp$}{i1}
   \fmflabel{$\cp$}{i2}
   \fmflabel{$\sqp'$}{i3}
   \fmflabel{$\sqp'$}{i4}
   \fmflabel{$v_{64}$}{j1}
   \fmfdot{j1}
\end{fmfgraph*}

& \\ & & &  & &

\poi \\ $W_{42}(Q_{12}; t, t') \,\; \sqp \cp \; \sqp' \cp'$

& \hspace{0.5cm} & \hspace{0.5cm}

$W_{43}(Q_{12}; t, t') \,\; \sqp \cp \; \sqp' \sqp'$

& &

$W_{43}(Q_{12}; t, t') \,\; \sqp \cp \; \sqp' \sqp'$

& \\ & {\footnotesize (4)}& & {\footnotesize (5)}& &
{\footnotesize (6)}\\ \hline & & &  & & \\

\hspace{0.5cm}
\begin{fmfgraph*}(3.9,1.9)
   \fmfleftn{i}{2}
   \fmfrightn{o}{2}
   \fmf{plain}{i1,v1}
   \fmf{dashes}{i2,v1}
   \fmf{phantom}{o1,v2}
   \fmf{phantom}{o2,v2}
   \fmf{phantom,left,tag=1,tension=1/2}{v1,v2}
   \fmf{phantom,left,tag=1,tension=1/2}{v2,v1}
   \fmffreeze
   \fmfi{plain}{subpath (0,1) of vpath1(__v1,__v2)}
   \fmfi{dashes}{subpath (1,2) of vpath1(__v1,__v2)}
   \fmfi{plain}{subpath (0,1) of vpath1(__v2,__v1)}
   \fmfi{plain}{subpath (2,1) of vpath1(__v2,__v1)}
   \fmfdot{v1,v2}
   \fmflabel{$V_{41}$}{v1}
   \fmflabel{$V_{21}$}{v2}
   \fmflabel{$\cp$}{i1}
   \fmflabel{$\sqp$}{i2}
\end{fmfgraph*}

& \hspace{0.5cm} &

\hspace{0.5cm}
\begin{fmfgraph*}(3.9,1.9)
   \fmfleftn{i}{2}
   \fmfrightn{o}{2}
   \fmf{plain}{i1,v1}
   \fmf{dashes}{i2,v1}
   \fmf{phantom}{o1,v2}
   \fmf{phantom}{o2,v2}
   \fmf{phantom,left,tag=1,tension=1/2}{v1,v2}
   \fmf{phantom,left,tag=1,tension=1/2}{v2,v1}
   \fmffreeze
   \fmfi{plain}{subpath (0,1) of vpath1(__v1,__v2)}
   \fmfi{plain}{subpath (1,2) of vpath1(__v1,__v2)}
   \fmfi{plain}{subpath (0,1) of vpath1(__v2,__v1)}
   \fmfi{plain}{subpath (2,1) of vpath1(__v2,__v1)}
   \fmfdot{v1}
   \fmflabel{$W_{41}$}{v1}
   \fmflabel{$\cp$}{i1}
   \fmflabel{$\sqp$}{i2}
\end{fmfgraph*}

& &

\hspace{0.5cm}
\begin{fmfgraph*}(3.4,1.5)
   \fmfbottomn{i}{2}
   \fmf{dashes,label=$$,label.side=left}{i1,v1}
   \fmf{plain,label=$$,label.side=left}{v1,i2}
   \fmf{plain,tag=1,tension=1/2}{v1,v1}
   \fmffreeze
   \fmfdot{v1}
   \fmflabel{$W_{41}$}{v1}
   \fmflabel{$\sqp$}{i1}
   \fmflabel{$\cp'$}{i2}
\end{fmfgraph*}

& \\ & & &  & &

\poi \\ $V_{21}(t) \,\; \sqp  \cp$

& \hspace{0.5cm} & \hspace{0.5cm}

$V_{21}(t) \,\; \sqp  \cp$

& &

$W_{21}(k; t, t') \,\; \sqp \; \cp'$

& \\ & {\footnotesize (7)}& & {\footnotesize (8)}& &
{\footnotesize (9)}\\ \hline  & & &  & & \\ & & &  & & \\

\hspace{0.5cm}
\begin{fmfgraph*}(3.4,1.3)
   \fmfbottomn{i}{2}
   \fmf{plain}{v1,i2}
   \fmf{dashes}{i1,v1}
   \fmf{phantom,tag=1,tension=1/2}{v1,v1}
   \fmffreeze
   \fmfi{dashes}{subpath (0,1) of vpath1(__v1,__v1)}
   \fmfi{plain}{subpath (1,2) of vpath1(__v1,__v1)}
   \fmfdot{v1}
   \fmflabel{$W_{42}$}{v1}
   \fmflabel{$\sqp$}{i1}
   \fmflabel{$\cp'$}{i2}
\end{fmfgraph*}

& \hspace{0.5cm} & \hspace{0.5cm}

\begin{fmfgraph*}(3.4,1.3)
   \fmfbottomn{i}{2}
   \fmf{dashes}{v1,i2}
   \fmf{dashes}{i1,v1}
   \fmf{phantom,tag=1,tension=1/2}{v1,v1}
   \fmffreeze
   \fmfi{plain}{subpath (0,1) of vpath1(__v1,__v1)}
   \fmfi{plain}{subpath (1,2) of vpath1(__v1,__v1)}
   \fmfdot{v1}
   \fmflabel{$W_{42}$}{v1}
   \fmflabel{$\sqp$}{i1}
   \fmflabel{$\sqp'$}{i2}
\end{fmfgraph*}

& &

\hspace{0.5cm}
\begin{fmfgraph*}(3.4,1.3)
\fmfbottomn{i}{2}
   \fmf{dashes}{v1,i2}
   \fmf{dashes}{i1,v1}
   \fmf{phantom,tag=1,tension=1/2}{v1,v1}
   \fmffreeze
   \fmfi{dashes}{subpath (0,1) of vpath1(__v1,__v1)}
   \fmfi{plain}{subpath (1,2) of vpath1(__v1,__v1)}
   \fmfdot{v1}
   \fmflabel{$W_{43}$}{v1}
   \fmflabel{$\sqp$}{i1}
   \fmflabel{$\sqp'$}{i2}
\end{fmfgraph*}&
\\ & & &  & &

\poi \\ $W_{21}(k; t, t') \,\; \sqp \; \cp'$

& \hspace{0.5cm} & \hspace{0.5cm}

$W_{22}(k; t, t') \,\; \sqp \; \sqp'$

& &

$W_{22}(k; t, t') \,\; \sqp \; \sqp'$

& \\ & {\footnotesize $\!\!\!\!(10)\!$}& & {\footnotesize
$\!\!\!\!\!(11)\!$}& & {\footnotesize $\!\!\!\!(12)\!$} \\ \hline

\end{tabular}
\end{fmffile}
\end{center}

\vspace{1mm}

{\footnotesize \noi FIG. 10.  Set of diagrams obtained after
performing two infinitesimal steps of RG transformation starting
from the initial condition given by Eq. (\ref{the action 1}). A
third step does not create new couplings.}


\poi Some conclusions can be drawn. If $m^2$, $\kappa$ and $\nu$
are initially zero, $m^2$ will be of order $\lambda$ through
$V_{41}$ in diagram 1 of Fig. 9, and $\kappa$ and $\nu$ will be at
least of order $\lambda^2$, through the one vertex diagrams of
Fig. 10. Note, as well, that $\eta$ will be of order $\lambda^2$
owing to the fact that $W_{21}$ is order $\lambda^2$ and is the
only term of the type $\sqp \cp$ which depends on $k$. This means
that, at order $\lambda^2$, $\eta$ can be omitted and $\alpha_t$
can be set equal to 1 in the left-hand side members of Eq.
(\ref{de vnm}), and Eqs. (\ref{de m2})-(\ref{de nu}). We will
assume that this is the case.

The vertices which depend on two times, have half of the fields
evaluated a $t$ and half at $t'$. Diagrams which depend on two
times with an uneven number of fields at $t$ and $t'$ can be drawn
using a $v_{6i}$ vertex, but they are zero. This fact can be
understood noting that, after $N$ iterations of the RG
transformation, $v_{6i}(k; t, t')$  will be given by a sum of $N$
terms, with the nth term proportional to $\delta(k - b^{-n})$
(remember that $\Lambda = 1$). So, if $k$ is not in the interval
$[b^{-1}, b^{-N}]$, $v_{6i}(k; t, t')$ is zero, or in other words
(and it will shown below) \bea v_{6i}(k; t, t') \propto \Theta(k -
1^+) \Theta(e^{s} 1^+ - k), \label{delta modulo}\ena where we use
$1^+$ meaning that $\Theta(k - 1^+)$ is strictly zero if $k = 1$.
Then, consider for example the case of \bea
v_{61}(|\bfk_1+\bfk_2+\bfk_3|; t, t') \; \sqp \cp \cp \; \cp' \cp'
\cp'. \ena If we connect the two $\cp$ fields to generate a term
of the form \bea \sqp \; \cp' \cp' \cp',\ena we must set $\bfk_2 =
-\bfk_3 = 1$ and integrate over $\hat\Omega$. But because of
(\ref{delta modulo}), the unpaired $\sqp$ field should have
momentum \bea |\bfk_1| > 1,\ena which is impossible, because
$\bfk_1 \le 1$. The same occurs if we connect two of the $\cp'$.
The conservation delta, $\delta^d(\bfk_1+ \bfk_2 \dots)$, which is
always present, allows us to write the first condition on Eq.
(\ref{delta modulo}) as \bea \Theta(|\bfk_4+\bfk_5+\bfk_6|-1^+).
\ena Setting, for example, $\bfk_5 = -\bfk_6 = 1$, it should be
$|\bfk_4|> 1$, and the conclusion follows.

There are other possible couplings depending on $t$ and $t'$, with
an even number of fields at each time, that due to causality are
not generated to order $\lambda^2$. For example, from \bea
v^{(2)}_{63} \; \sqp \sqp \sqp \; \cp' \cp' \cp' \ena joining one
of the $\sqp$ with one of the $\cp'$ we cannot generate a new term
of the form \bea \sqp \sqp \; \cp' \cp', \ena because, on one
hand, necessarily is $t' \le t$ (see Sec. \ref{section the
action}), and on the other hand, the propagator $\mG(t',t)$
connecting $\cp'$ with $\sqp$ entails the opposite condition $t'
\ge t$. Also, $v_{65} \; \sqp \sqp \sqp \; \sqp' \sqp' \cp'$
cannot create a term of the form $\sqp \sqp \; \sqp' \sqp'$,
because by construction (but not necessarily, as in the previous
example) is $t'\le t$.

Other typical cancellation occurs in the diagram of Fig. 11
\begin{center}
\vspace{0.9cm}
\begin{fmffile}{diag7}
\begin{fmfgraph*}(3.6,2.1)
   \fmfleftn{i}{2}
   \fmfrightn{o}{2}
   \fmf{dashes}{i1,v1}
   \fmf{dashes}{i2,v1}
   \fmf{phantom}{o1,v2}
   \fmf{phantom}{o2,v2}
   \fmf{phantom,left,label=$\propto\Theta(t-t')$,tag=1,tension=1/2}{v1,v2}
   \fmf{phantom,left,label=$\propto\Theta(t'-t)$,tag=1,tension=1/2}{v2,v1}
   \fmffreeze
   \fmfi{plain}{subpath (0,1) of vpath1(__v1,__v2)}
   \fmfi{dashes}{subpath (1,2) of vpath1(__v1,__v2)}
   \fmfi{plain}{subpath (0,1) of vpath1(__v2,__v1)}
   \fmfi{dashes}{subpath (2,1) of vpath1(__v2,__v1)}
   \fmfdot{v1,v2}
   \fmflabel{$V_{43}$}{v1}
   \fmflabel{$V_{21}\;\;\;\;\;\; = \;\;\;0$}{v2}
   \fmflabel{$\sqp$}{i1}
   \fmflabel{$\sqp$}{i2}
\end{fmfgraph*}
\end{fmffile}
\end{center}
{\footnotesize FIG. 11. Example of a diagram which is null due to
a closed chain of steps functions of the form: $\Theta(t_1-t_2)
\Theta(t_2 - t_3) \dots \Theta(t_{n-1} - t_n) \Theta(t_n-t_1)$.
This is typical of diagrams drawn using vertices which are local
in time.}

\poi There are other cases that can be analyzed in similar terms.

\subsection{Final remark}

We have seen in Sec. \ref{obtaining delta S} the method used to
extract from diagrams with two external lines the terms which by
definition belong to the free action, Eq. (\ref{the action 2}).
Something analogous must be done with diagrams which give terms
with 4 external fields of the form \bea \sqp \cp \cp' \cp' \ena
and \bea \sqp \cp \sqp' \sqp'. \ena For example, the contributions
of diagrams (2) and (5) of Fig. 10 can be written as \bea
\label{diagram W41} \D\int_{0}^{T} \!\! dt \D\int_{0}^{T}\!\!
dt'\!\! \D\int_{b\Lambda} \!\!{d^d\!k_1 \dots d^d\!k_4} \,
\delta^d\lt(\bfk_1+\dots\rt) \, \sqp(\bfk_1, t) \cp(\bfk_2, t) \;
\cp(\bfk_3, t') \cp(\bfk_4, t')\; v(Q_{12}; t, t') \delta s, \ena
\bea \label{diagram W43} \D\int_{0}^{T} \!\! dt \D\int_{0}^{T}
\!\! dt'\!\! \D\int_{b\Lambda}\!\! {d^d\!k_1 \dots d^d\!k_4} \,
\delta^d\lt(\bfk_1+\dots \rt) \, \sqp(\bfk_1, t) \cp(\bfk_2, t) \;
\sqp(\bfk_3, t') \sqp(\bfk_4, t')\; w(Q_{12}; t, t') \delta s,
\ena where \bea v(k; t, t') = 18 \, \delta_{k; 0} \; V_{41}^2 \;
\mG\cG(t, t'), \ena and  \bea w(k; t, t') = 18 \, \delta_{k; 0} \;
V_{41} V_{43} \; \mG\cG(t, t'). \ena In principle, $v\: \delta s$
and $w \: \delta s$ are not directly identified with $\delta
W_{41}$ and $\delta W_{43}$, but can contain contributions to
$\delta V_{41}$ and $\delta V_{43}$. Using the definition
(\ref{def cP}), we write \bea v(k; t, t') \, \delta s &=& \delta(t
- t')\; \delta V_{41} + \delta W_{41}(k; t, t'), \\ \nn \\ w(k; t,
t') \, \delta s &=& \delta(t - t')\; \delta V_{43} + \delta
W_{43}(k; t, t') \ena with \bea \delta V_{41} = \cP v(0) \, \delta
s, \ena and so for $V_{43}$. In this way,  the corrections to
$V_{41}$ and $V_{43}$ are isolated, while $\delta W_{41}$ and
$\delta W_{43}$  give the net contributions to the $W_{41}$ and
$W_{43}$.

\section{Detailed expressions for $\kappa$ and $\nu$ \label{appendix:detailed
expressions}}

Here, the formulas for $\kappa$ and $\nu$, with $\beta_0^{-1} = 0$
are given. The expressions are for $k = 0$, but are asymptotically
valid for $0\le k$ when $1 \ll z$. They are composed by two parts,
corresponding to $z < 2$ and $z  > 2$, which join smoothly for
$z\ra 2$.

\bea \kappa(z, T) = \kappa_<(z, T) \; \Theta(2-z) + \kappa_>(z, T)
\; \Theta(z-2), \ena \bea \nu(z, T) = \nu_<(z, T) \; \Theta(2-z) +
\nu_>(z, T) \; \Theta(z-2). \ena

\subsection{Expressions for $\kappa$}

We write \bea \kappa_< = V_0^2 \D\sum_{i=1}^7 \kappa_{<i}\;, \ena
and an analogous equation for $\kappa_>$, where

\bea \kappa_{<1} =  \D\frac{3}{2T} \Big\{- 63 z \, \log 3
-\D\frac{4}{z} (3 - 4 z + z^2) + 6 (-5 + 3 z) \cos T - 14 (-1 + z)
\cos(3 T)\Big\},\ena

\bea  \kappa_{<2} = \D\frac{18}{2T} \Big\{- \cos\lt[T (-2 + z)\rt]
+ \D\frac{2}{z} \cos(T z) -  7 \cos\lt[T (2 + z)\rt] \nn \\ +
\cos\lt[T (1 - 2 z)\rt] + 7 \cos\lt[T (1 + 2  z)\rt] \Big\}, \ena

\bea   \kappa_{<3} = \D\frac{9}{2T} \Big\{z \, \log(-2 + z) - 2 z
(5 + 4 \cos T ) \, \log z  - 7 z \log(2 + z) \nn \\ - 4 z \log(-1
+ 2 z) + 28 z \log(1 + 2 z)\Big\}, \ena

\bea   \kappa_{<4} = -18 \lt[z \sin T -  \sin(T z)\rt], \ena

\bea \kappa_{<5} =  3\Big\{-3\lt(1 + 3 z \rt) \,\Si(T) - 21 (-1 +
z) \,\Si(3 T) - 3(-2 + z) \,\Si[T (-2 + z)] \nn \\ +  12 \,\Si(T
z)  - 21 (2 + z) \,\Si[T (2 + z)] +  3 (-1 + 2 z) \,\Si[T(-1 + 2
z)] \nn \\ + 21 (1 + 2 z) \,\Si[T (1 + 2 z)]\Big\}, \ena

\bea \kappa_{<6} =  18 z T \lt[\Ci(T) - \Ci(T z)\rt] + \D\frac{9z
}{2T}\Big\{-3 \Ci(T) -  \Ci[T(-2 + z)] + 7  \Ci[T (2 + z)] \nn \\
+ 21 \Ci(3 T) + 4 \Ci[T (-1 + 2 z)] - 28  \Ci[T (1+ 2 z)]\Big\},
\ena

\bea \kappa_{<7} =  \D\frac{9 z}{T}\Big\{- 12 \big[\Ci(2 T)  -
\Ci(2 T z)\big]
\cos T +  3  \lt[\Ci(T)  - \Ci(T z)\rt] \cos(2 T) \nn \\
16 \lt[\Si(2T) - \Si(2 T z)\rt] \sin T  - 4 \lt[\Si(T) - \Si(T
z)\rt] \sin(2T) \Big\}, \ena

\bea \kappa_{>1} =  -\D\frac{201}{8 T} - 9 T + \D\frac{63}{8 T z}
+ \D\frac{3z}{32 T} \Big\{97 \nn \\ - 24  \big[\gamma_E + 21 \log
3 + \log 64 + \log T - (2 + \log 4) T^2 \big] \Big\}, \ena

\bea  \kappa_{>2} =  \D\frac{3}{32T z} \bigg(z^2 \Big[144 \cos T -
24 (1+\log 4) \cos(2 T) - 112 \cos(3 T) - 105 \cos(4 T)\Big] \nn
\\ - 28 \cos(2 T z)  +  z \Big\{-240 \cos T + 120 \cos(2 T)  + 112
\cos(3 T) - 84 \cos(4 T)  \nn \\  - 24 \cos[2 T (-1 + z)] + 48
\cos[T(1 - 2 z)] + 336 \cos[T(1 + 2 z)]\Big\}\bigg),\ena

\bea \kappa_{>3} =   -\D\frac{9}{8} \Big\{z \big[8 \sin T - 4
\sin(2 T) - 7 \sin(4 T)\big] + 14 \sin(2 T z)\Big\},  \ena

\bea \kappa_{>4} =  -\D\frac{9 z}{4T} \big[2 (9 + T^2) + 8 \cos T
- \cos(2 T)\big] \log z  \nn \\ - \D\frac{9 z}{4T}  \Big[- \log(-1
+ z) + 4 \log(-1 + 2 z) - 28 \log(1 + 2 z)\Big],  \ena

\bea \kappa_{>5} =  \D\frac{9 z}{4T} \Big\{(-3 + 4 T^2) \Ci(T) +
\Ci(2 T) + 21 \Ci(3 T) + 7 \Ci(4 T) - \Ci[2 T (-1 + z)] \nn \\ - 2
T^2 \Big[2 \Ci(2 T) + 7 \Ci(4 T) - 7 \Ci(2 T z)\Big] + 4 \Ci[T (-1
+ 2 z)] - 28 \Ci[T (1 + 2 z)]\Big\}, \ena

\bea \kappa_{>6} =  \D\frac{9}{2} \Big\{-(1 + 3 z) \Si(T) + (3 +
z) \Si(2 T) - 7 (-1 + z) \Si(3 T) - 7 (1 + z) \Si(4 T) \nn \\  +
(1 - z) \Si[2 T (-1 + z)] - 7 \Si(2 T z) + (-1 + 2 z) \Si[T (-1 +
2 z)] \nn \\ + 7 (1 + 2 z) \Si[T(1 + 2 z)]\Big\},\ena

\bea \kappa_{>7} =   \D\frac{9 z}{8T} \Big\{\Big[12 \Ci(T) - 12
\Ci(2 T) - 2 \Ci(4 T) + 2 \Ci(2 T z) \Big] \cos(2 T) \nn \\ + 64
\lt[\Si(2 T) - \Si(2 T z)\rt] \sin T  -  2 \lt[8 \Si(T) - 8 \Si(2
T) + \Si(4 T) - \Si(2 T z)\rt] \sin(2 T) \nn \\ + 48 \lt[-\Ci(2 T)
+ \Ci(2 T z)\rt] \cos T \Big\}. \ena The functions $\Si$ and $\Ci$
are the $\sin$ and $\cos$ integral functions, and $\gamma_E$ is
the Euler's constant.

\subsection{Expressions for $\nu$}

We write \bea \nu_< = V_0^2 \D\sum_{i=1}^7 \nu_{<i}\;, \ena and an
analogous equation for $\nu_>$, where

\bea \nu_{<1} = \D\frac{9}{4T} \lt[-12 -\log 9 - 32 z + \lt(44 -
\log 2187\rt) z^2 \rt]),\ena

\bea \nu_{<2} =  27 z \lt[-z \sin T  + \sin(T z) \rt], \ena

\bea \nu_{<3} =  \D\frac{9}{2T} \Big\{-3 (1 + 7 z^2) \cos T  - (-1
+ z^2) \cos(3 T) + (-2 + z) \cos[T (-2 + z)] \nn \\ + 18 \cos(T z)
- (2 + z) \cos[T (2 + z)] +7 (-1 + 2 z) \cos[T(1 - 2  z)] \nn \\ +
(1 + 2 z) \cos[T(1 + 2 z)]\Big\}, \ena

\bea  \nu_{<4} = \D\frac92 \Big\{-3 (1 + 9 z^2) \, \Si(T) - 3 (-1
+ z^2) \, \Si(3 T) + (-2 + z)^2 \, \Si[T (-2 + z)]\nn \\ + 24 z \,
\Si(T z) - (2 + z)^2 \, \Si[T (2 + z)] - 7 (1 - 2 z)^2 \, \Si[T
(1- 2 z)] \nn \\ + (1 + 2 z)^2 \, \Si[T(1 + 2 z)]\Big\}, \ena

\bea  \nu_{<5} =  \D\D\frac9{4T} \Big\{3 \lt[-2 + (-19 + 4 T^2)
z^2\rt] \, \Ci(T) + (2 + 7 z^2) \, \Ci(3 T) \nn \\ + (-4 + z^2) \,
\Ci[T (-2 + z)]   + 12(2 - T^2 z^2) \, \Ci(T z) + (-4 + z^2) \,
\Ci[T (2 + z)] \nn \\  + 14(-1 + 4 z^2) \, \Ci[T (-1 + 2 z)]   +
2(1 - 4 z^2) \, \Ci[T(1 + 2 z)] \Big\},\ena

\bea   \nu_{<6} =  \D\frac9{4T} \Big\{(4 - z^2) \log(-4 + z^2) -
2(12 -25 z^2) \log z \nn \\  -  2(-1 + 4 z^2) \lt[7\log(-1 + 2 z)
- \log(1 + 2 z)\rt] \Big\}, \ena

\bea  \nu_{<7} =  \D\frac{9 z^2}{2T} \Big\{\!24 \lt[\Ci(2 T) -
\Ci(2 T z)\rt] \cos T +  \lt[\Ci(T) - \Ci(T z)\rt] \cos(2 T) \nn
\\ + 32 \lt[\Si(2 T) - \Si(2 T z)\rt] \sin T \Big\}, \ena

\bea  \nu_{>1} =  \D\frac{9}{16T} \Big\{-4 \lt(7 + 4 \gamma_E + 2
\log 1536 + 4 \log T \rt) - 56 z \nn \\ + \lt[135 + 4 \gamma_E +
28 \log(4/3) + 4 \log T \rt] z^2 \Big\},\ena

\bea  \nu_{>2} =  \D\frac94 \Big\{- 2 z \sin(2 T z) - \lt[12 \sin
T - 6 \sin(2 T) - \sin(4 T)\rt]z^2  \Big\},\ena

\bea \nu_{>3} =  \D\frac9{16T} \Big\{-24 (1 + 7 z^2) \cos T + 4(14
- 4 z + 11 z^2) \cos (2 T) \nn \\ - 8 (-1 + z^2) \cos (3 T)  -  (8
+ 3 z^2) \cos (4 T) - 56 (-1 + z) \cos [2 T (-1 + z)] \nn \\  - 12
\cos (2 T z)  + 56 (-1 + 2 z) \cos [T (1 - 2 z)] + 8 (1 + 2 z)
\cos [T (1+ 2 z)] \Big\}, \ena

\bea \nu_{>4} =  \D\frac{9}{4T}\Big\{3 \lt[-2 - (19 - 4 T^2)
z^2\rt] \Ci(T) + 2 \lt[5 - (-7 + 6 T^2) z^2\rt] \Ci(2 T) \nn \\ +
(2 + 7 z^2) \Ci(3 T) - (2 -z^2 + 4 T^2 z^2) \Ci(4 T) + 14 (1 -
z^2) \Ci[2 T (-1 + z)] \nn
\\ - (2 - 4 T^2 z^2) \Ci(2 T z) - 14(1 - 4 z^2) \Ci[T (-1 + 2
z)] + (2 - 8 z^2) \Ci[T (1+ 2 z)]\Big\}, \ena

\bea \nu_{>5} =  \D\frac{9}{2} \Big\{\!-3 (1 + 9 z^2) \Si(T) + 2
(7 - 2 z + 7 z^2) \Si(2 T) + 3(1 -  z^2) \Si(3 T) \nn \\ + (-4 -
z^2) \Si(4 T) - 14(1 - 2 z + z^2) \Si[2T(-1 + z)] - 4 z \Si(2 T z)
\nn
\\ - 7 (1 - 4 z + 4 z^2) \Si[T(1 - 2 z)] + (1 + 4 z + 4 z^2)
\Si[T(1 + 2 z)] \Big\}, \ena

\bea \nu_{>6} =  \D\frac{9}{2T} \Big\{7(-1 + z^2) \log(-1 + z)  +
(1 + 17 z^2) \log z  \nn \\ -  (-1 + 4 z^2) \lt[7\log(-1 + 2 z) -
\log(1 + 2 z)\rt]\Big\}, \ena

\bea \nu_{>7} =  \D\frac{9z^2}{2T} \Big\{24 \lt[\Ci(2 T) - \Ci(2 T
z)\rt] \cos T \nn \\ + \lt[\Ci(T) - \Ci(2 T) - 7 \Ci(4 T) + 7
\Ci(2 T z)\rt] \cos(2 T) \nn \\ + 32 \lt[\Si(2 T) - \Si(2 T z)\rt]
\sin T + 7 \lt[-\Si(4 T) + \Si(2 T z)\rt] \sin(2 T) \Big\}. \ena

\end{document}